\documentclass[11pt, secnumarabic,amssymb, nobibnotes, aps, prd]{revtex4-2}
	
\usepackage[dvips]{graphicx,color}
\usepackage{times}
\usepackage{filecontents}
\usepackage{mathtools}
\usepackage{subcaption}
\usepackage{graphicx}
\usepackage{amsmath}
\usepackage{multirow}
\usepackage{xcolor}

\usepackage[%
  colorlinks=true,
  urlcolor=blue,
  linkcolor=red,
  citecolor=blue
]{hyperref}

\begin{document}
\title{Thermodynamic topology of 4D Dyonic AdS black holes in different ensembles}

\author{Naba Jyoti Gogoi$^1$  }\email{gogoin799@gmail.com }
\author{Prabwal Phukon$^1$} \email{prabwal@dibru.ac.in}

\affiliation{$1.$ Department of Physics, Dibrugarh University,
Dibrugarh 786004, Assam, India}

\begin{abstract}
We study the thermodynamic topology of four dimensional dyonic Anti-de-Sitter(AdS) black hole in three different ensembles: canonical, mixed and grand canonical ensemble. While canonical ensemble refers to the ensemble with fixed electric and magnetic charges, mixed ensemble is an ensemble where we fix magnetic charge and electric potential. In the grand canonical ensemble, potentials corresponding to both electric and magnetic charges are kept fixed. In each of these ensembles, we first compute the topological charges associated with critical points. We find that while in both canonical and mixed ensembles, there exists one conventional critical point with topological charge $-1$, in the grand canonical ensemble, we find no critical point. Then, we consider the dyonic AdS black hole as topological defects in thermodynamic space and study its local and global topology by computing the winding numbers at the defects. We observe that while the topologies of the black hole in canonical and mixed ensembles are identical with total topological charge equaling $1$, in the grand canonical ensemble, depending on the values of potentials, the total topological charge is either equal to $0$ or $1$. In canonical and mixed ensembles, either one generation and one annihilation points or no generation/annihilation points are found. In the grand canonical ensemble, depending on the values of potentials, we find either one generation point or no generation/annihilation point. Thus, we infer that the topological class of $4$D dyonic AdS black hole is ensemble dependent.
\end{abstract}


\maketitle
\section{Introduction}
Thermodynamic phase behavior of black holes has been studied extensively since the early days of black hole thermodynamics \cite{Hawking, Bekenstein73, Hawking:1974sw, Bekenstein:1972tm, Bekenstein:1973ur, Bardeen:1973gs, Phy, bekenstein1980black, Wald:1999vt, Carlip:2014pma, Wall:2018ydq, Candelas:1977zz, Chamblin:1999hg, Hawking:1982dh, Chamblin:1999tk}. In recent years, a lot of focus has been attributed to the study of criticality in AdS black holes in extended thermodynamic space  \cite{Kubiznak:2012wp,Altamirano:2013ane,Altamirano:2013uqa,Wei:2014hba,Frassino:2014pha,Cai:2013qga,Xu:2014tja,Dolan:2014vba,Hennigar:2015esa,Hennigar:2015wxa,Hennigar:2016xwd,Zou:2016sab},  where the cosmological constant $\Lambda$ is considered as thermodynamic pressure $P$ \cite{Kastor:2009wy,Dolan:2012jh, Gunasekaran:2012dq,Chen:2016gzz} .
\begin{equation}
P=-\frac{\Lambda}{8 \pi G}
\end{equation}
where, $G$ is Newton's  gravitational constant. Accordingly,  a thermodynamic volume $V$ is defined conjugate to thermodynamic pressure $P$ and the first law of black hole thermodynamics takes the revised form: 
\begin{equation}
dM= T dS + V dP+ \sum_i Y_i dx^i,
\end{equation}
where, $T$ is the temperature, $S$ is the entropy and  $Y_i dx^i$ is the $i$-th chemical potential term.

A recent addition to the study of criticality in black holes is the the idea of thermodynamic topology. Initiated in \cite{Wei:2021vdx}, in this novel approach,  Duan’s topological current $\phi$-mapping theory \cite{Duan} is invoked in the thermodynamic space of a black hole to study its criticality. Consequently, the critical points in the thermodynamic space are characterized with distinct topological charges and based on those charges, are classified into conventional and novel critical points. The key steps involved are summarized below : 

The temperature, $T$, of a black hole is expressed as a function of pressure $P$, entropy $S$ and other thermodynamic parameters.
\begin{equation}
T=T(S,P,x^i),
\end{equation}
where $x^i$ denotes other thermodynamic parameters. Then, pressure is, then, eliminated by using  $(\partial_S T)_{P,x^i}=0,$ and a new potential $\Phi$, known as Duan's potential  is constructed.
\begin{equation}
\Phi=\frac{1}{\sin \theta} T(S,x^i).
\end{equation}
The term  `$1/\sin \theta$' is introduced for the sake of convenience. A two dimensional vector  $\phi=(\phi^\theta,\phi^\theta)$ is defined following the framework of Duan's $\phi$-mapping theory \cite{Duan,Duan:2018rbd} as
\begin{equation}
\phi^{S}=(\partial_{S} \Phi)_{\theta,x^i}  \quad , \quad \phi^\theta=(\partial_{\theta} \Phi)_{S,x^i}.
\end{equation}
The presence of $\theta$ in the vector field $\phi$  ensures that the zero point of the vector field $\phi$ is always at $\theta=\pi/2$. The critical points can be calculated using this criteria. Also the topological current, $j^\mu$, is conserved i.e.,  $\phi^a(x^i)=0$. This construction ensures the existence of topological charge which for a given parameter region $\Sigma$ is equal to
\begin{equation}
Q=\int_\Sigma j^0 d^2x=\sum_{i=1} ^N \beta_i n_i=\sum_{i=1} ^N  w_i.
\end{equation}
Here,  $w_i$, $j^0$ and $\beta_i$ are the winding number of $i$-th zero points of $\phi$, the density of the topological current and the Hopf index  respectively. Critical points with topological charges $-1$ and $+1$ are referred as conventional critical point and novel critical point respectively. The total topological charge of a black hole is computed as the sum of individual charges associated with each critical point. Following the work in  \cite{Wei:2021vdx}, analysis of thermodynamic topology has been extended to a number of black holes \cite{Yerra:2022alz,Bai:2022klw,Yerra:2022eov,Wei:2022mzv,Yerra:2022coh,Ahmed:2022kyv,Bai:2022vmx}.\\

An alternative way to apply topology in black hole thermodynamics has also been proposed in \cite{Wei:2022dzw}. In this method,  black hole solutions are regarded as defects in thermodynamic parameter spaces. These defects are then studied in terms of their winding numbers. The sign of the winding number of a defect has been linked to the thermodynamic stability of the corresponding black hole solution.  The sum of the winding numbers is, now, termed as topological number based on which different black hole solutions are categorized. The analysis begins with the introduction of a generalized free energy $F$ defined as follows: 
\begin{equation}
\label{Generalized_Free_Energy}
\mathcal{F}=E-\frac{S}{\tau},
\end{equation}
where $E$ and $S$ are the energy, entropy. $\tau$ is a quantity which has the dimension of time. A vector field $\phi$ is defined  from $F$ in the following way :  
\begin{equation}
\label{Eq:Topological_Defect_Vector_Field}
\phi=\Big( \frac{\partial\mathcal{F}}{\partial r_+}, -\cot\Theta \csc\Theta \Big)
\end{equation}
The zero point of the vector $\phi$ is at $\Theta=\pi/2$.
The unit vector is defined as 
\begin{equation}
\label{Eq:TD_Unit_Vector}
n^a=\frac{\phi^a}{||\phi||} \quad (a=1,2) \quad \text{and} \quad \phi^1=\phi^{r_+}, \quad \phi^2=\phi^\Theta. 
\end{equation}
Corresponding to a given value of $\tau$, the zero points of  $n^1$ are computed.The winding numbers of each of these zero points are calculated. The topological number of a black hole is obtained by summing over the individual winding numbers of all the black hole branches. Following \cite{Wei:2022dzw}, study of black holes as topological defects has been extended to a number of black holes in \cite{Wu:2022whe,Liu:2022aqt,Fan:2022bsq,Fang:2022rsb,Ye:2023gmk,Zhang:2023uay,Du:2023wwg,Sharqui:2023mbx,Du:2023nkr,Wu:2023sue,Wu:2023xpq}.\\

Motivated by all the above mentioned works, in this paper, we extend the study of thermodynamic topology to $4$d dyonic AdS black hole in different ensembles. Our primary focus is to understand the ensemble dependent nature of thermodynamic topology. For this, we carry out our analysis in three different ensembles: $1.$ canonical ensemble where both the electric and the magnetic charges are kept fixed, $2.$ mixed ensemble where electric potential and magnetic charge are kept fixed and $3.$ grand canonical ensemble where both electric and magnetic potentials are kept fixed. To begin with, in each of these ensembles, we figure out the critical points and compute their topological charges. Based on the sign of those charges, we classify them as conventional and novel critical points. Then we consider the black hole as a topological defect in each of these ensembles and find out its topological number,  generation and annihilation points in the thermodynamic space.\\

This paper is organized as follows. In \autoref{Section:Dyonic_Canonical}, we begin with $4$d dyonic AdS black hole in canonical ensemble and study its thermodynamic topology. This is followed by similar studies  in mixed ensemble in \autoref{Section:Dyonic_Mixed_Canonical} and grand canonical ensemble in \autoref{Section:Dyonic_Grand_Canonical}. We conclude with our findings in \autoref{Section:Canonical}.

\section{$4$d Dyonic AdS black hole in canonical ensemble}
\label{Section:Dyonic_Canonical}
The four dimensional, asymptotically anti-de Sitter, dyonic black holes  solution has its origin in maximal gauged supergravity . Such a black hole carries both electric and magnetic charges. The dyonic black hole solution can be obtained by the reduction of five dimensional Kaluza-Klein theory and it has some very interesting properties \cite{Dobiasch:1981vh,Gibbons:1994ff,Rasheed:1995zv,Campbell:1992hc,Cheng:1993wp,Lu:2013ura}. A simpler solution of dyonic AdS black hole can be obtained by varying the Reissner-Nordström action with a cosmological constant \cite{Dutta:2013dca}. The four dimensional, asymptotically anti de-Sitter, dyonic black hole metric is given by :
\begin{equation}
ds^2=-f(r)dt^2+\frac{1}{f(r)}dr^2+r^2d\theta^2+r^2\sin^2\theta d\phi^2,
\end{equation}
where,
\begin{equation}
f(r)=\frac{q_e^2+q_m^2}{r^2}+\frac{r^2}{l^2}-\frac{2 M}{r}+1,
\end{equation}
Here, $q_e$, $q_m$, $M$ and $l$ are electric charge, magnetic charge, mass of the black hole and the AdS radius respectively.
Thermodynamic pressure, $P$, is related to the AdS radius as 
\begin{equation}
P=\frac{3}{8 \pi  l^2}.
\end{equation}
The mass, $M$ and the entropy, $S$ of the black hole are given by (in the following expressions, $r_+$ denotes the horizon radius)
\begin{equation}
\label{Eq:Dyonic_Canonical_Mass}
	\begin{aligned}	
		M &=\frac{l^2 q_e^2+l^2 q_m^2+l^2 r_+^2+r_+^4}{2 l^2 r_+}=\frac{3 q_e^2+3 q_m^2+8 \pi  P r_+^4+3 r_+^2}{6 r_+} 
	\end{aligned}	
\end{equation}
\begin{equation}
\label{Eq:Entropy_Canonical}
S=\pi r_+^2,
\end{equation}
 
\subsection{Topology of $4$d dyonic AdS black hole thermodynamics in canonical ensemble}
\begin{figure}[h!]
	\centerline{
	\includegraphics[scale=0.7]{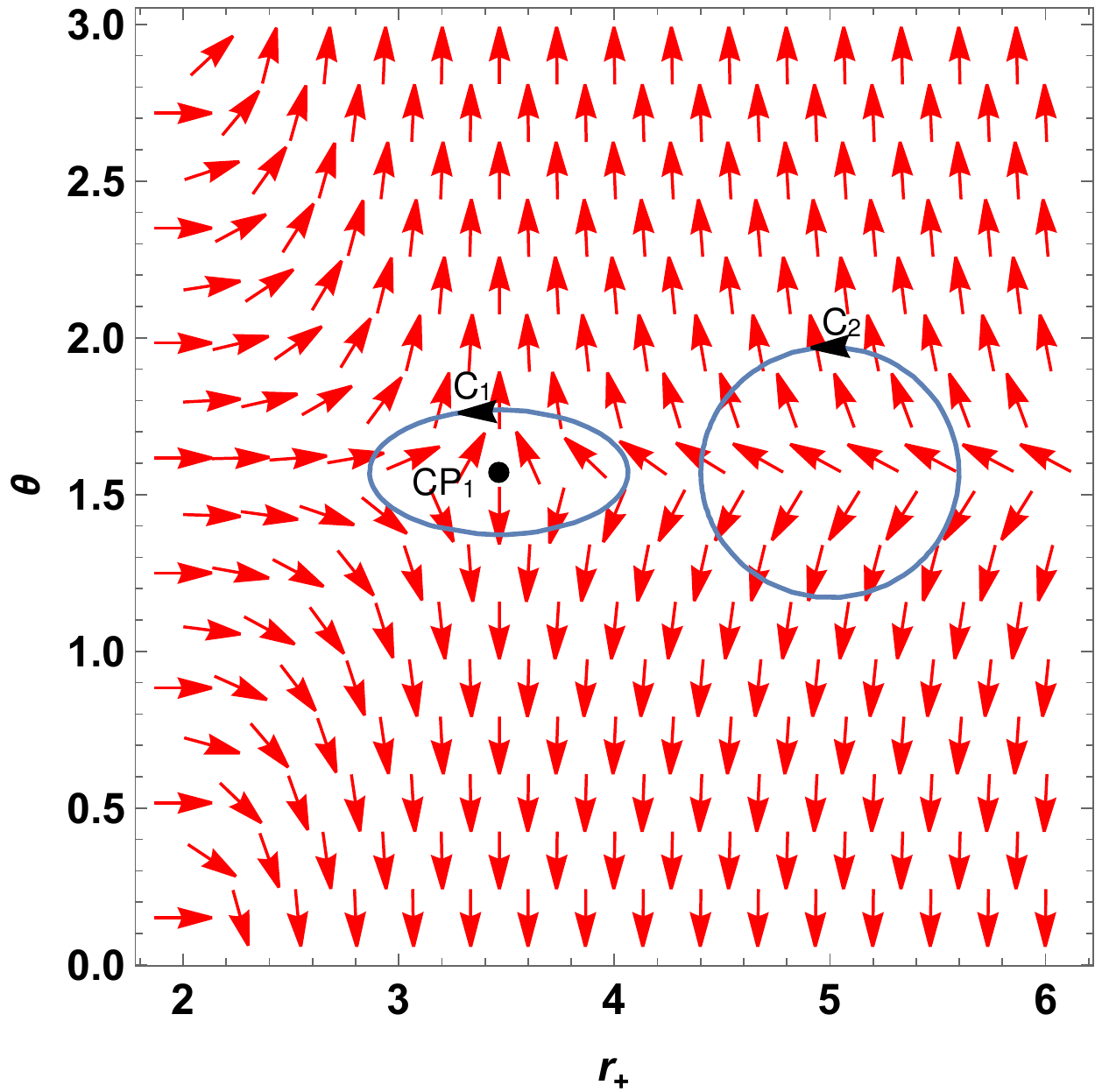}}
	\caption	{Plot of the normalized vector field $n$ in $r_+$ vs $\theta$ plane for dyonic AdS black hole in canonical ensemble. The black dot represents the critical point.}
	\label{Fig:Dyonic_Canonical_Topology_Vector_Field}
	\end{figure}

To study the topology of dyonic AdS black hole thermodynamics, we write the temperature as a function of pressure, horizon radius, electric and magnetic charges.
\begin{equation}
\label{Eq:Temperature_Canonical_General_with_P}
T=\frac{\partial_{r_+} M}{\partial_{r_+} S}=\frac{8 \pi  P r_+^4+r_+^2-q_e^2-q_m^2}{4 \pi  r_+^3}.
\end{equation}
Use of the condition $\Big( \frac{\partial_{r_+} T}{\partial_{r_+}S}\Big)_{q_e,q_m,P}=0$ leads us to an expression for pressure, $P$.
\begin{equation}
P=\frac{r_+^2-3 \left(q_e^2+q_m^2\right)}{8 \pi  r_+^4},
\end{equation}
Plugging $P$ in \eqref{Eq:Temperature_Canonical_General_with_P}, we get rid of the pressure term and the temperature, $T$ takes the following form :
\begin{equation}
\label{Eq:Temperature_Canonical}
T(q_e,q_m,r_+)=\frac{r_+^2-2 \left(q_e^2+q_m^2\right)}{2 \pi  r_+^3}.
\end{equation}
A thermodynamic function $\Phi$ is defined as,
\begin{equation}
	\begin{aligned}
		\Phi&=\frac{1}{\sin \theta}T(q_e,q_m,r_+) \\
		    &=\frac{\csc \theta  \left\{ r_+^2-2 \left(q_e^2+q_m^2\right)\right\}}{2 \pi  r_+^3},
	\end{aligned}
\end{equation}
The vector components of the vector field $\phi=( \phi^{r_+},\phi^\theta )$ are
\begin{equation}
\phi^{r_+}=\Big( \frac{\partial \Phi}{\partial r_+} \Big)_{q_e,q_m,\theta}=-\frac{\csc \theta  \left\{r_+^2-6 \left(q_e^2+q_m^2\right)\right\}}{2 \pi  r_+^4},
\end{equation}
and
\begin{equation}
\phi^\theta=\Big( \frac{\partial \Phi}{\partial \theta} \Big)_{q_e,q_m,r_+}=-\frac{\cot \theta  \csc \theta  \left\{r_+^2-2 \left(q_e^2+q_m^2\right)\right\}}{2 \pi  r_+^3}
\end{equation}
The normalized vector components are 
\begin{equation}
\label{Eq:Canonical_Normalizd_Vector_Component_1}
\frac{\phi^{r_+}}{||\phi||}=\frac{6 \left(q_e^2+q_m^2\right)-r_+^2}{\sqrt{r_+^2 \cot ^2(\theta ) \left(r_+^2-2 \left(q_e^2+q_m^2\right)\right){}^2+\left(r_+^2-6 \left(q_e^2+q_m^2\right)\right){}^2}}, 
\end{equation}
and
\begin{equation}
\frac{\phi^{\theta}}{||\phi||}=-\frac{r_+ \cot \theta  \left\{ r_+^2-2 \left(q_e^2+q_m^2\right)\right\}}{\sqrt{r_+^2 \cot ^2\theta  \left\{ r_+^2-2 \left(q_e^2+q_m^2\right)\right\}{}^2+\left\{ r_+^2-6 \left(q_e^2+q_m^2\right)\right\}{}^2}}
\end{equation}
The normalized vector  $n=\Big(\frac{\phi^{r_+}}{||\phi||},\frac{\phi^{\theta}}{||\phi||}\Big)$ has been plotted in \autoref{Fig:Dyonic_Canonical_Topology_Vector_Field}. This figure shows the vector plot of $n$ in a $r_+$ vs $\theta$ plane for dyonic AdS black hole. For this plot, we  have fixed $q_e=q_m=1$. The black dot represents the critical point ($CP_1$). To calculate the critical point we set $\theta=\pi/2$ in \eqref{Eq:Canonical_Normalizd_Vector_Component_1} and equate this to zero. The critical point is located at $(r_+,\theta)=(\sqrt{6(q_e^2+q_m^2)},\pi/2)$ or at $(r_+,\theta)=(2\sqrt{3},\pi/2)$ for $q_e=q_m=1$.

For the calculation of the topological charge of the critical point, a contour $C$ parametrized by $\vartheta\in(0,2\pi)$ is defined \cite{Wei:2021vdx} as follows:
\begin{equation}	
\label{Contour}
	\begin{cases}
		&r_+=a\cos\vartheta+r_0, \\
		&\theta=b\sin\vartheta+\frac{\pi}{2}.
	\end{cases}
\end{equation}
We construct two contours $C_1$ and $C_2$ where the first contour encloses the critical point $CP_1$ and the second contour is outside the critical point. For these contours we choose  $(a,b,r_0)=(0.6,0.2,2\sqrt{3})$ and $( 0.6,0.4,5)$.

The deflection  of the vector field $n$ along the contour $C$ is,
\begin{equation}
\label{Deflection Angle}
\Omega(\vartheta)=\int_0^\vartheta \epsilon_{ab}n^a\partial_\vartheta n^b  d\vartheta.
\end{equation}
The topological charge is, then, equal to,  $Q=\frac{1}{2\pi}\Omega(2\pi)$. For the critical point $CP_1$ enclosed by the contour $C_1$,  the topological charge has been found to be, $Q_{CP_1}=-1$. This is a conventional critical point. Since the contour $C_2$ does not enclose any critical point, it corresponds to zero topological charge. Thus, the total topological charge is, $Q=-1$.
\begin{figure}[h!]
	\centerline{
	\includegraphics[scale=0.7]{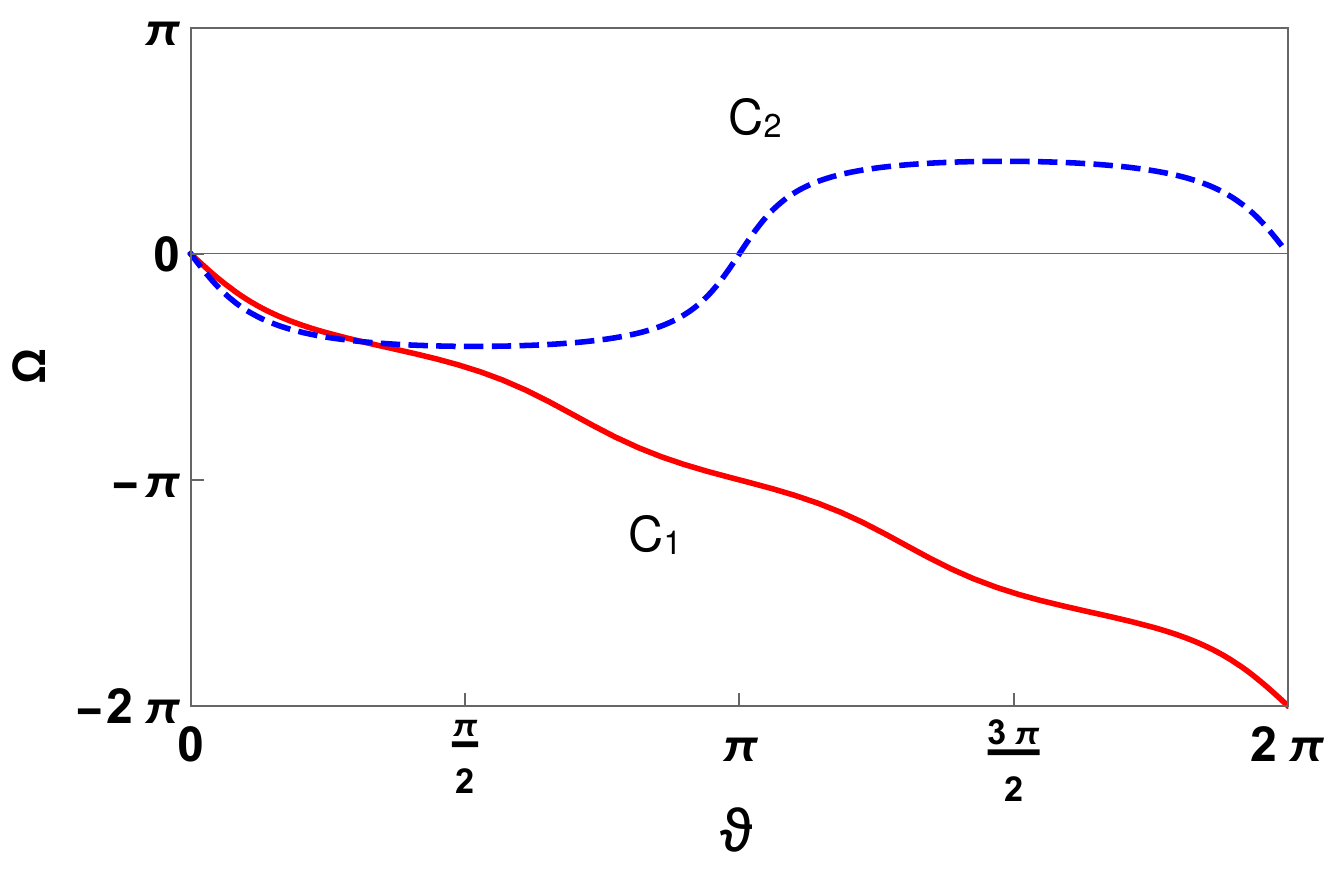}}
	\caption	{$\Omega$ vs $\vartheta$ plot for the contour $C1$ and $C_2$}
	\label{Fig:Dyonic_Canonical_Topology_Deflection_Angle}
	\end{figure}
The behaviour of $\Omega$ is shown in \autoref{Fig:Dyonic_Canonical_Topology_Deflection_Angle}. The red curve corresponds $C_1$ and the blue curve corresponds to $C_2$. The function $\Omega(\vartheta)$ for $C_1$ decreases non-linearly and reaches $-2\pi$ at $\vartheta=2\pi$. On the other hand, $\Omega(\vartheta)$ reaches zero at $\vartheta=2\pi$ for $C_2$.

$4d$ dyonic AdS black hole in canonical ensemble has the following equation of state : 
\begin{equation}
T=\frac{8 \pi  P r_+^4+r_+^2-q_e^2-q_m^2}{4 \pi  r_+^3},
\end{equation}
The corresponding critical points are given by,
\begin{equation}
\label{Eq:Dyonic_Canonical_Critical_Values}
T_c=\frac{1}{3 \sqrt{6} \pi  \sqrt{q_e^2+q_m^2}}, \quad P_c=\frac{1}{96 \pi  q_e^2+96 \pi  q_m^2} \quad \text{and} \quad r_c= \sqrt{6(q_e^2+q_m^2)}
\end{equation}

\begin{figure}[h!]
	\centerline{
	\includegraphics[scale=0.7]{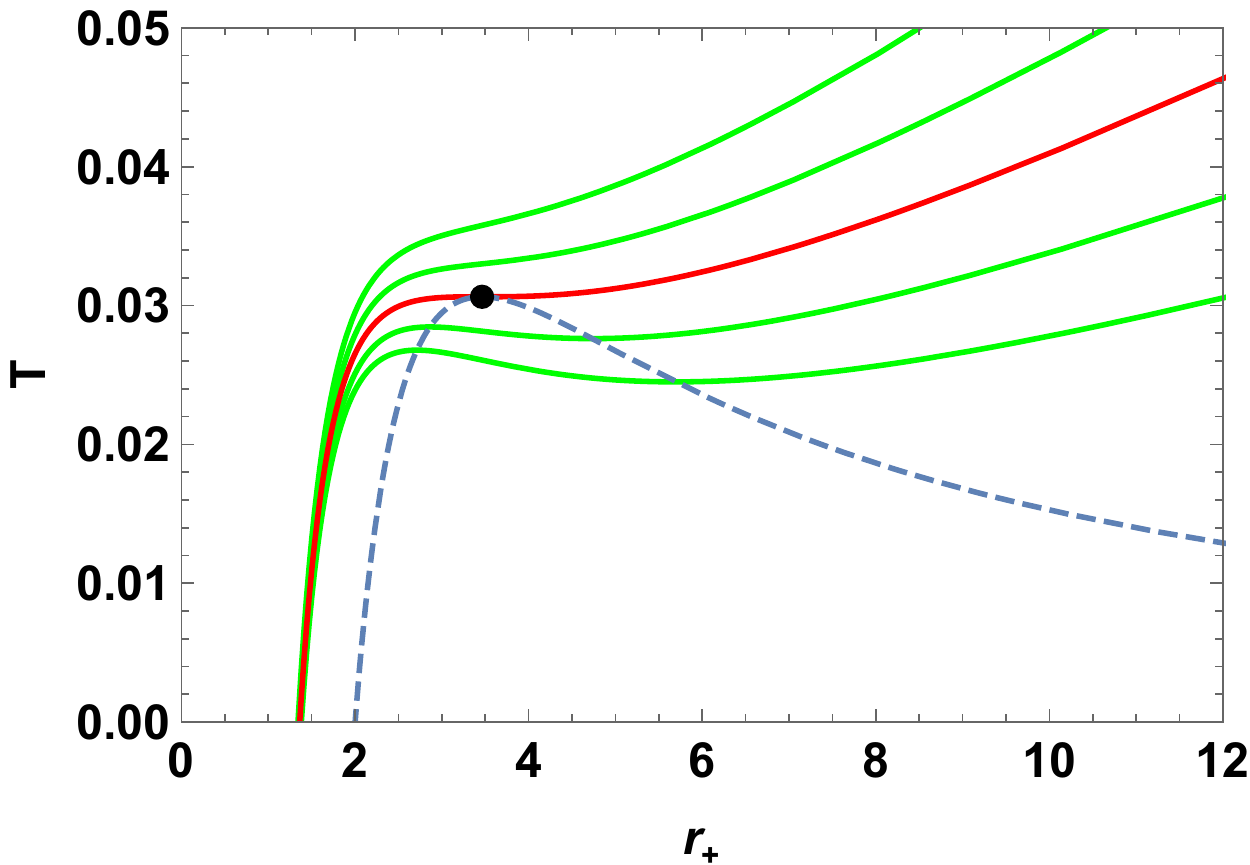}}
	\caption	{Isobaric curves (red and green) of dyonic AdS black hole in canonical ensemble. Black dot represents the critical point.}
	\label{Fig:Dyonic_Canonical_Critical_Point_Isobaric_Curve}
	\end{figure}

It can be clearly seen that the critical radius in \eqref{Eq:Dyonic_Canonical_Critical_Values} exactly matches the critical point obtained from thermodynamic topology,  $(r_+,\theta)=(\sqrt{6(q_e^2+q_m^2)},\pi/2)$.  As mentioned above this is a conventional critical point with topological charge $-1$. To see the nature of the critical point we plot the phase structure (isobaric curves) around it  in \autoref{Fig:Dyonic_Canonical_Critical_Point_Isobaric_Curve}. The location of the critical point in the isobaric curve is shown as a black dot.
The red curve is the isobaric curve for $P=P_c$. The green curves above and below the red curve are respectively for $P>P_c$ and $P<P_c$. The blue dashed curvet describes the extremal points and is plotted using \eqref{Eq:Temperature_Canonical}. From \autoref{Fig:Dyonic_Canonical_Critical_Point_Isobaric_Curve}, it is observed that for $P<P_c$, the small and large black hole phases are separated by the unstable region (the negative slope region of the isobaric curves or the region enclosed by the two extremal points corresponding to each isobaric curve). Different phases of the dyonic AdS black hole in canonical ensemble disappear at the critical point. Hence, the critical point $CP_1$ can be thought of as a phase annihilation point. 

\subsection{Dyonic AdS black hole solution as topological thermodynamic defects in canonical ensemble}
\begin{figure}[!ht]
	\centering
		\begin{subfigure}{0.5\textwidth}
			\centering
			\includegraphics[width=0.9\linewidth]{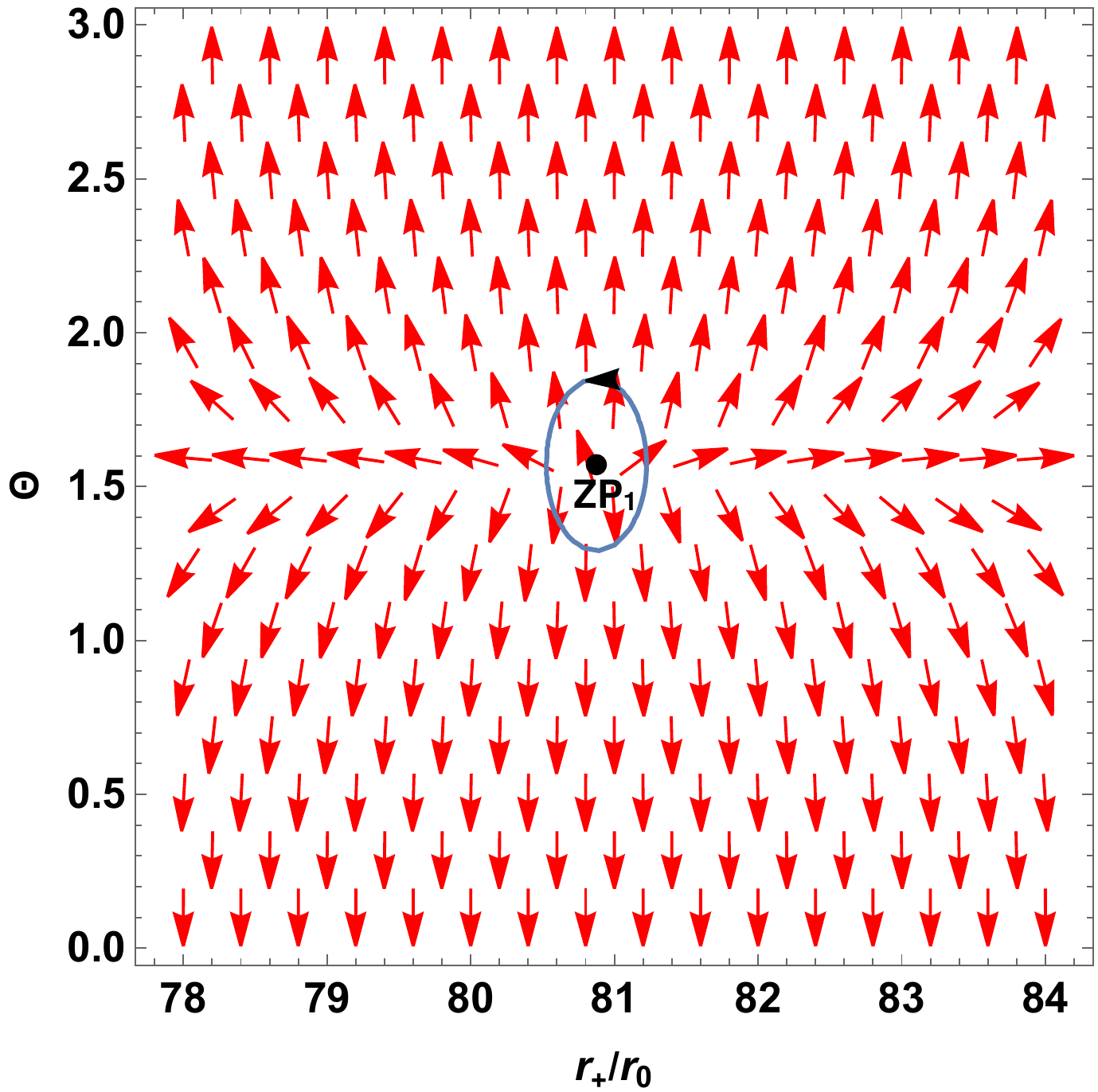}
			\caption{Unit vector $n$ in $\Theta$ vs $r_+/r_0$ plane for $\tau/r_0=30$}
			\label{Fig:Dyonic_Canonical_Topologycal_Defect_Vector_Plot_Left_Region}
		\end{subfigure}%
		\begin{subfigure}{0.5\textwidth}
			\centering
			\includegraphics[width=0.9\linewidth]{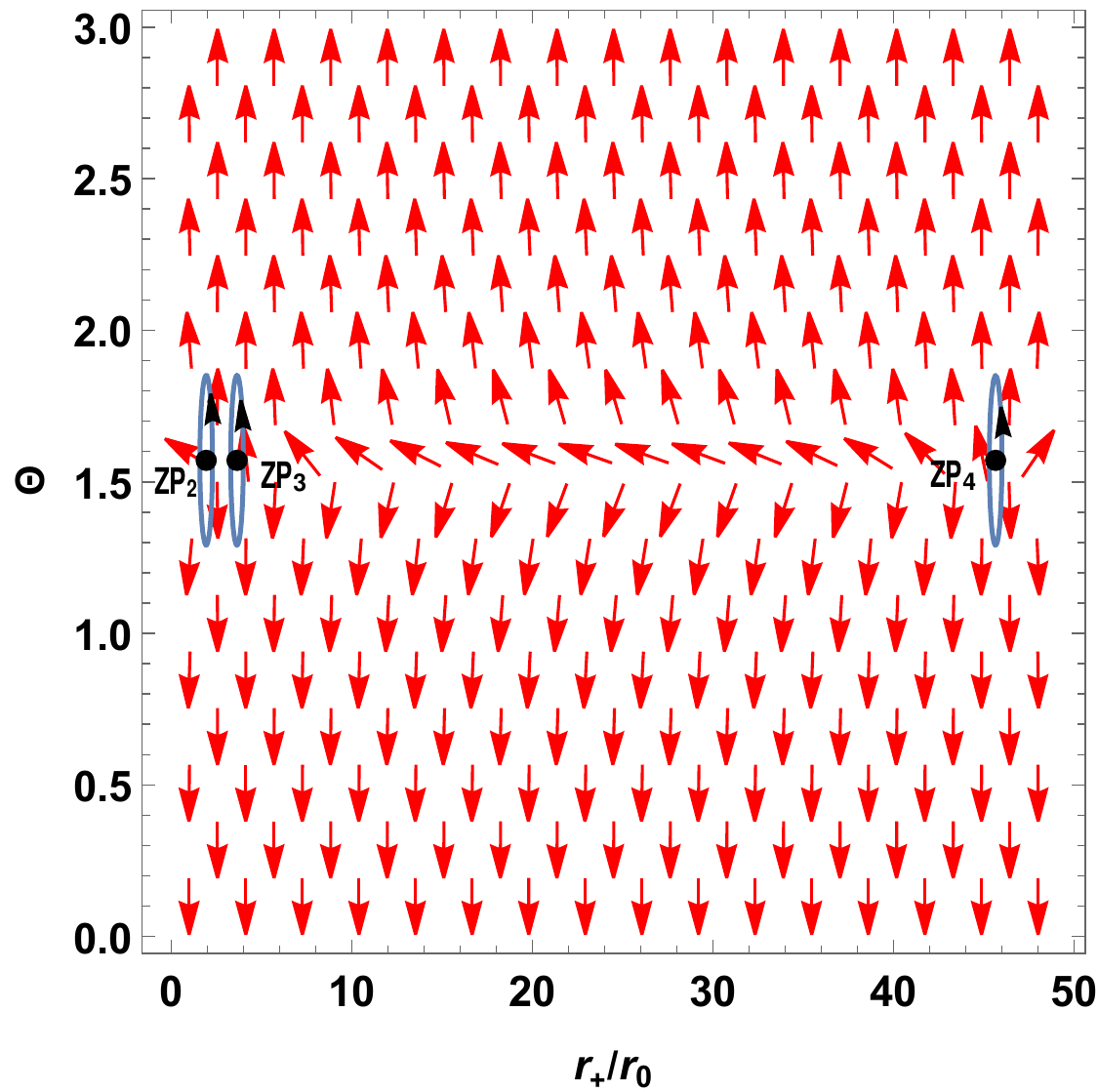}
			\caption{Unit vector $n$ in $\Theta$ vs $r_+/r_0$ plane for $\tau/r_0=50$. The zero points are $ZP_2$, $ZP_3$ and $ZP_4$ from left to right.}
			\label{Fig:Dyonic_Canonical_Topologycal_Defect_Vector_Plot_Middle_Region}
		\end{subfigure} \\
		\begin{subfigure}{0.5\textwidth}
			\centering
			\includegraphics[width=0.9\linewidth]{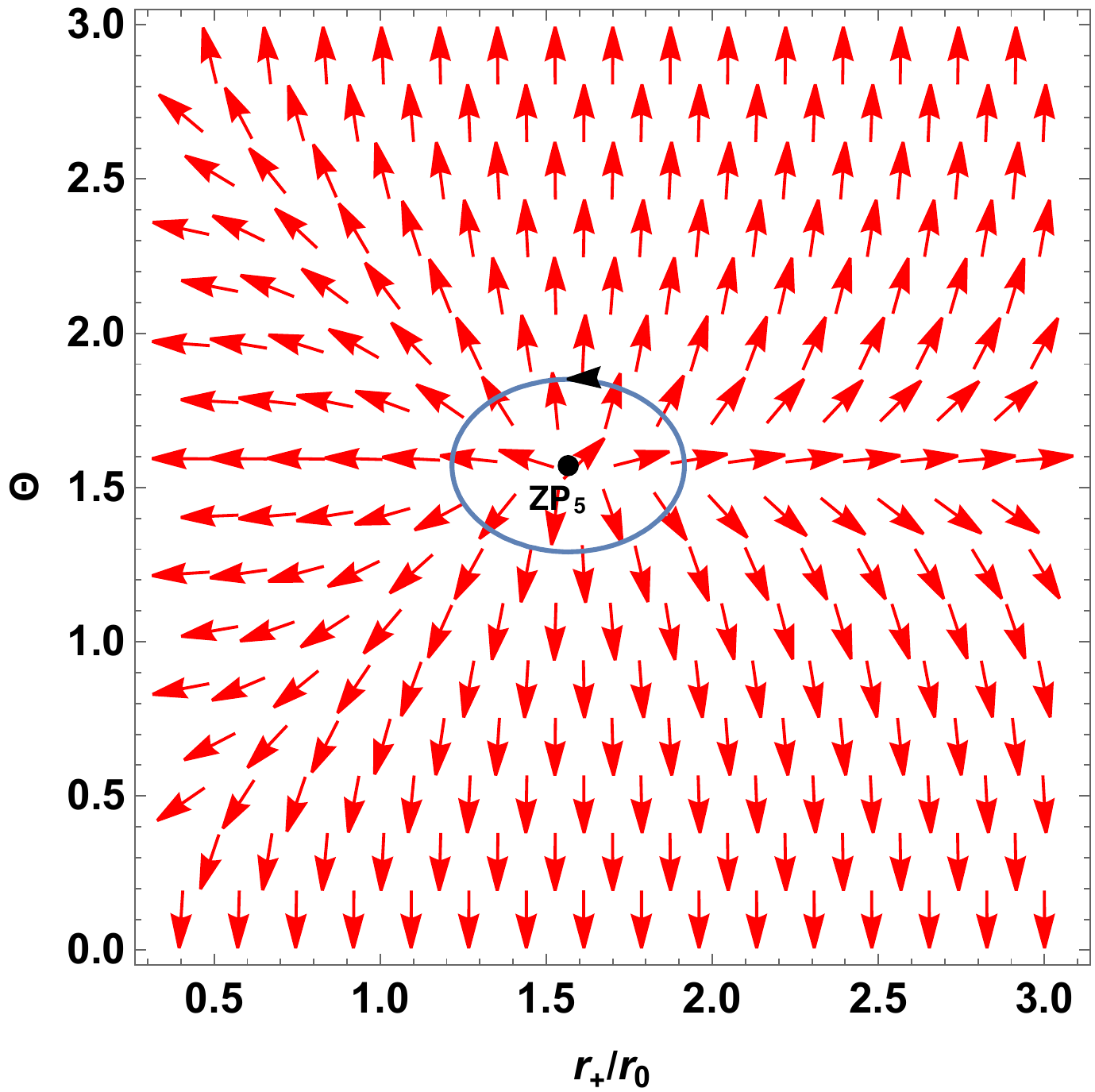}
			\caption{Unit vector $n$ in $\Theta$ vs $r_+/r_0$ plane for $\tau/r_0=100$}
			\label{Fig:Dyonic_Canonical_Topologycal_Defect_Vector_Plot_Right_Region}
		\end{subfigure}%
	\caption{Unit vector $n=(n^1,n^2)$ shown in $\Theta$ vs $r_+/r_0$ plane for $Pr_0^2=0.0002$   (below critical pressure $P_c$). The black dots represent the zero points.}
	\label{Fig:Dyonic_Canonical_Topologycal_Defect_Vector_Plot_All_Regions}
	\end{figure}

Now, we proceed to study the dyonic AdS black hole solution in canonical ensemble as topological thermodynamic defects. Using the mass and entropy of the black hole from \eqref{Eq:Dyonic_Canonical_Mass} and \eqref{Eq:Entropy_Canonical} in \eqref{Generalized_Free_Energy}, the generalized free energy is found to be,
\begin{equation}
\mathcal{F}=\frac{3 q_e^2+3 q_m^2+8 \pi  P r_+^4+3 r_+^2}{6 r_+}-\frac{\pi  r_+^2}{\tau }.
\end{equation}
The vector field components of the vector given by \eqref{Eq:Topological_Defect_Vector_Field} are
\begin{equation}
\label{Eq:Canonical_TD_First_Vector_Component}
\phi^{r_+}=\frac{1}{2}-\frac{q_e^2+q_m^2}{2 r_+^2}+4 \pi  P r_+^2-\frac{2 \pi  r_+}{\tau },
\end{equation}
and 
\begin{equation}
\phi^\Theta=-\cot \Theta  \csc \Theta .
\end{equation}
 The corresponding unit vectors are
\begin{equation}
\label{Eq:First_Unit_Vector_Canonical}
n^1=\frac{4 \pi  r_+^3 \left(2 P r_+ \tau -1\right)-\tau  \left(q_e^2+q_m^2-r_+^2\right)}{r_+^2 \tau  \sqrt{\frac{\left\{\tau  \left(q_e^2+q_m^2-r_+^2\right)+4 \pi  r_+^3 \left(1-2 P r_+ \tau \right)\right\} {}^2}{r_+^4 \tau ^2}+4 \cot ^2\Theta  \csc ^2 \Theta }}, 
\end{equation}
and 
\begin{equation}
n^2=-\frac{\cot \Theta  \csc \Theta }{\sqrt{\frac{\left\{ \tau  \left(q_e^2+q_m^2-r_+^2\right)+4 \pi  r_+^3 \left(1-2 P r_+ \tau \right)\right\}{}^2}{4 r_+^4 \tau ^2}+\cot ^2\Theta  \csc ^2\Theta }}.
\end{equation}
These unit vectors are plotted and used to locate the zero points by setting $\Theta=\pi/2$ in $n^1$ (see \eqref{Eq:First_Unit_Vector_Canonical}) 
and equating it to zero. For example, setting $q_e/r_0=1$, $q_m/r_0=1$, $Pr_0^2=0.0002$ and $\tau/r_0=30$, we can find one zero 
point ($ZP_1$) located at $(r_+/r_0,\Theta)=(80.8742,\pi/2)$. Here, $r_0$ is an arbitrary length scale which is determined by the size of a
 cavity that surrounds the black hole. The value of pressure is taken below the critical pressure $P_c$. The representation of the unit vectors 
along with the zero point is shown in \autoref{Fig:Dyonic_Canonical_Topologycal_Defect_Vector_Plot_Left_Region}. The winding number or
 the topological charge corresponding to this zero point is computed following the prescription stated in the previous section and found to 
be $w=+1$. Similarly, keeping the same charge and pressure configuration,  corresponding to $\tau/r_0=50$, we find three zero 
points $ZP_2$, $ZP_3$ and $ZP_4$ with winding numbers $+1$, $-1$ and $+1$ respectively. These are
 shown in \autoref{Fig:Dyonic_Canonical_Topologycal_Defect_Vector_Plot_Middle_Region}. 
For $\tau/r_0=100$, a solitary zero point $ZP_5$ with winding number $+1$ is observed (\autoref{Fig:Dyonic_Canonical_Topologycal_Defect_Vector_Plot_Right_Region}).

\begin{figure}[h]
	\centerline{
	\includegraphics[scale=0.7]{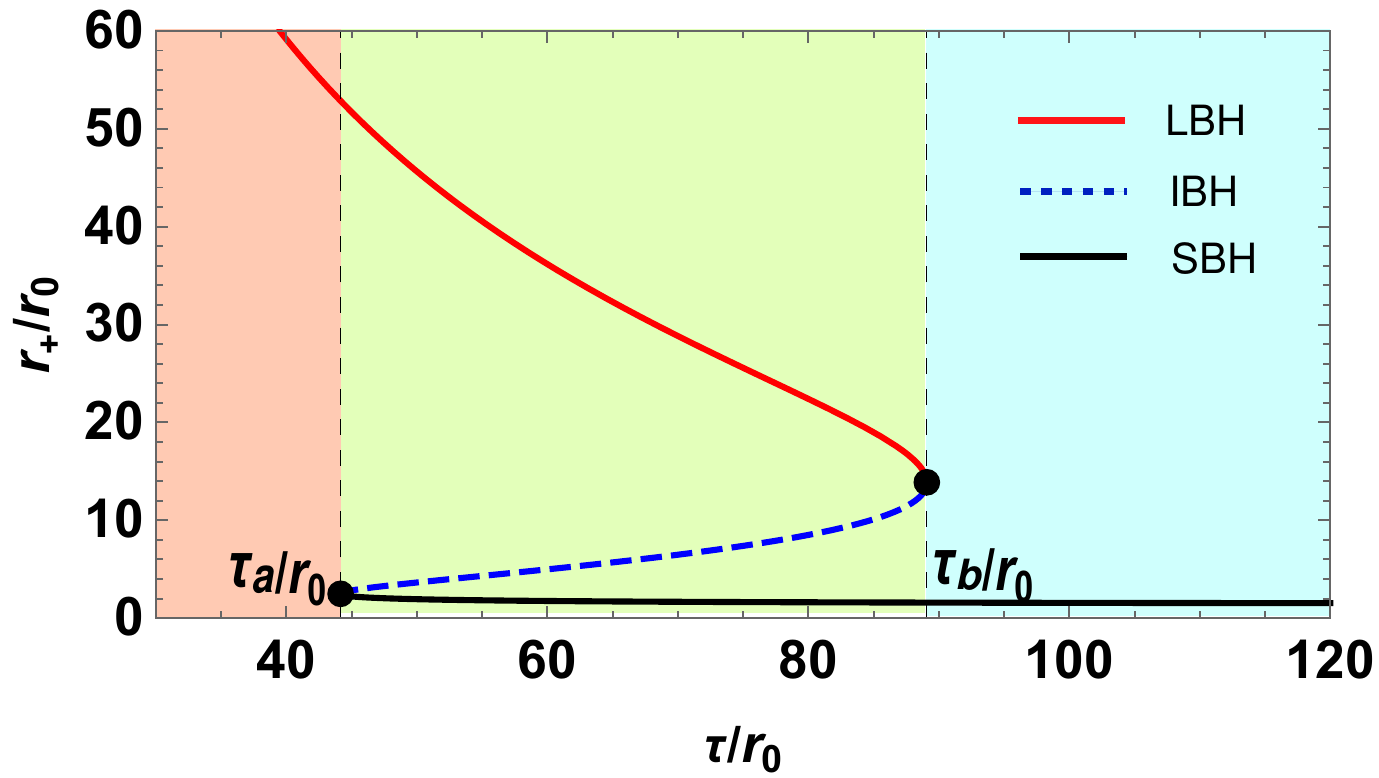}}
	\caption	{ The zero points of $\phi^{r_+}$ in $\tau/r_0$ vs $r_+/r_0$ plane for dyonic AdS black hole in canonical ensemble for pressure less than the critical pressure $P_c$.}
	\label{Fig:Dyonic_Canonical_Topologycal_Defect_Tau_Curve}
	\end{figure}

An analytic expression for $\tau$  corresponding to zero points can be otained by setting $\phi^{r_+}=0$. 
\begin{equation}
\tau=\frac{4 \pi  r_+^3}{8 \pi  P r_+^4+r_+^2-q_e^2-q_m^2}.
\end{equation}
A plot of $r_+$ vs $\tau$ obtained above is shown in \autoref{Fig:Dyonic_Canonical_Topologycal_Defect_Tau_Curve}. The points on this curve are the zero points of  $\phi^{r_+}$. Here, we have fixed  $q_e/r_0=1$, $q_m/r_0=1$ and  $Pr_0^2=0.0002$ (below the critical pressure $P_c$).\\

In \autoref{Fig:Dyonic_Canonical_Topologycal_Defect_Tau_Curve}, three different black hole branches are clearly visible. The branch $\tau<\tau_b$ corresponds to the large black hole region. The winding number for any zero point on this branch is found to be $w=+1$. Similarly, winding number $w=+1$ is also observed for any zero point on the branch $\tau>\tau_a$ which corresponds to the small black hole region. The branch $\tau_a<\tau<\tau_b$ represents the intermediate black hole region and winding number for any zero point on this branch is equal to $w=-1$. The topological number is, hence, $W=+1-1+1=+1$. We explicitly computed the specific heats at the three branches and found that the branches with winding number $+1$ have positive specific heat (thermodynamically stable) and the branch with winding number $-1$ has negative specific heat(thermodynamically unstable).\\

Finally, we find the generation/annihilation points  by using the condition $\partial_{r_+}\mathcal{F}=\partial_{r_+,r_+}\mathcal{F}=0$. For $q_e/r_0=1$, $q_m/r_0=1$, $Pr_0^2=0.0002$, we get the generation and annihilation points at $\tau/r_0=\tau_a/r_0=44.1585$ and $\tau/r_0=\tau_b/r_0=89.0811$  respectively which are shown as black dots in \autoref{Fig:Dyonic_Canonical_Topologycal_Defect_Tau_Curve}. \\

\begin{figure}[h]
	\centering
		\begin{subfigure}{0.5\textwidth}
			\centering
			\includegraphics[width=0.8\linewidth]{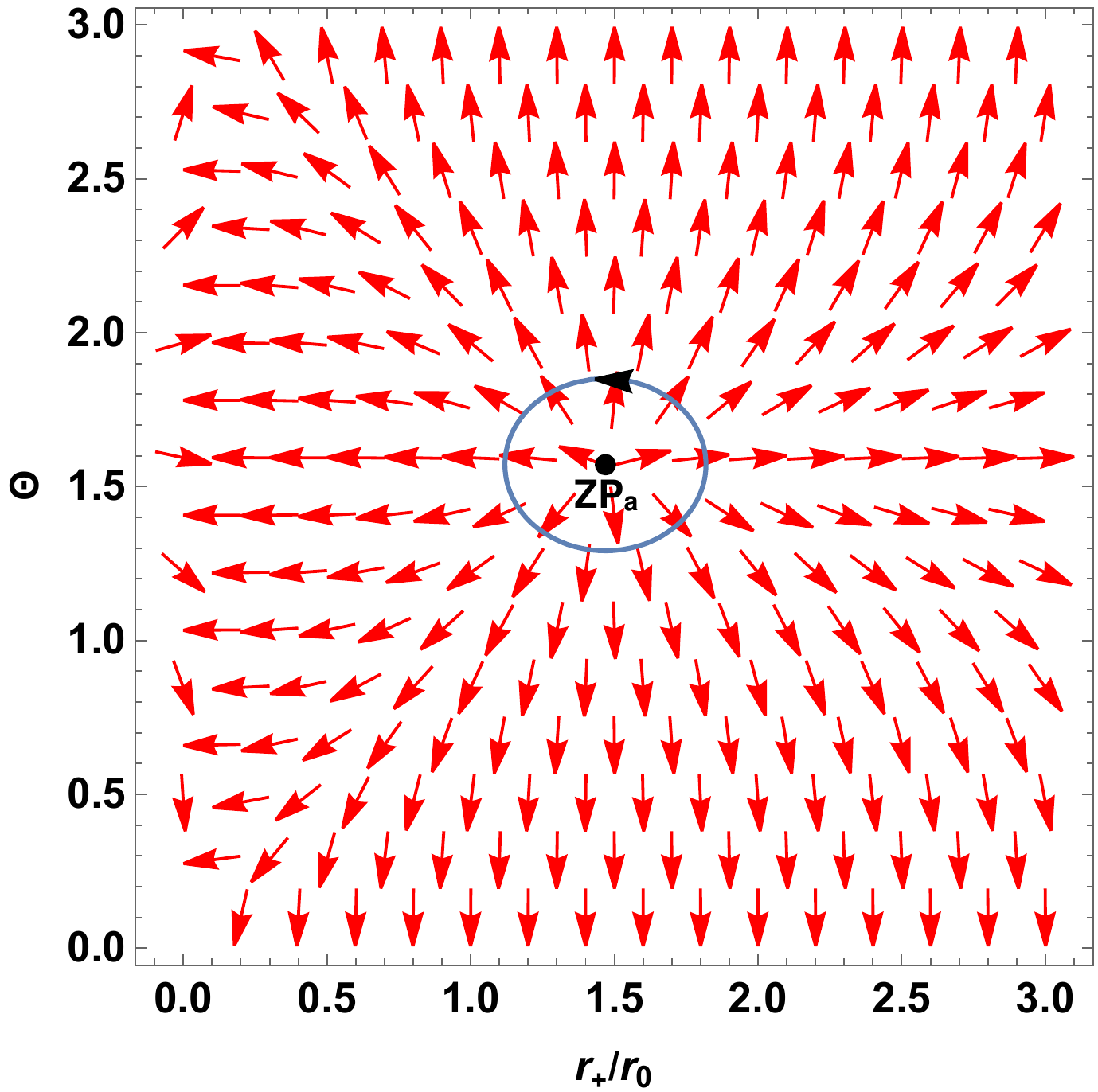}
			\caption{Unit vector $n=(n^1,n^2)$ shown in $\Theta$ vs $r_+/r_0$ plane for $Pr_0^2=0.01$. The black dot represents zero point.}
			\label{Fig:Dyonic_Canonical_Thermodynamic_Defect_Zero_Point_Vector_Above_Critical_Pressure}
		\end{subfigure}%
		\begin{subfigure}{0.5\textwidth}
			\centering
			\includegraphics[width=1.0\linewidth]{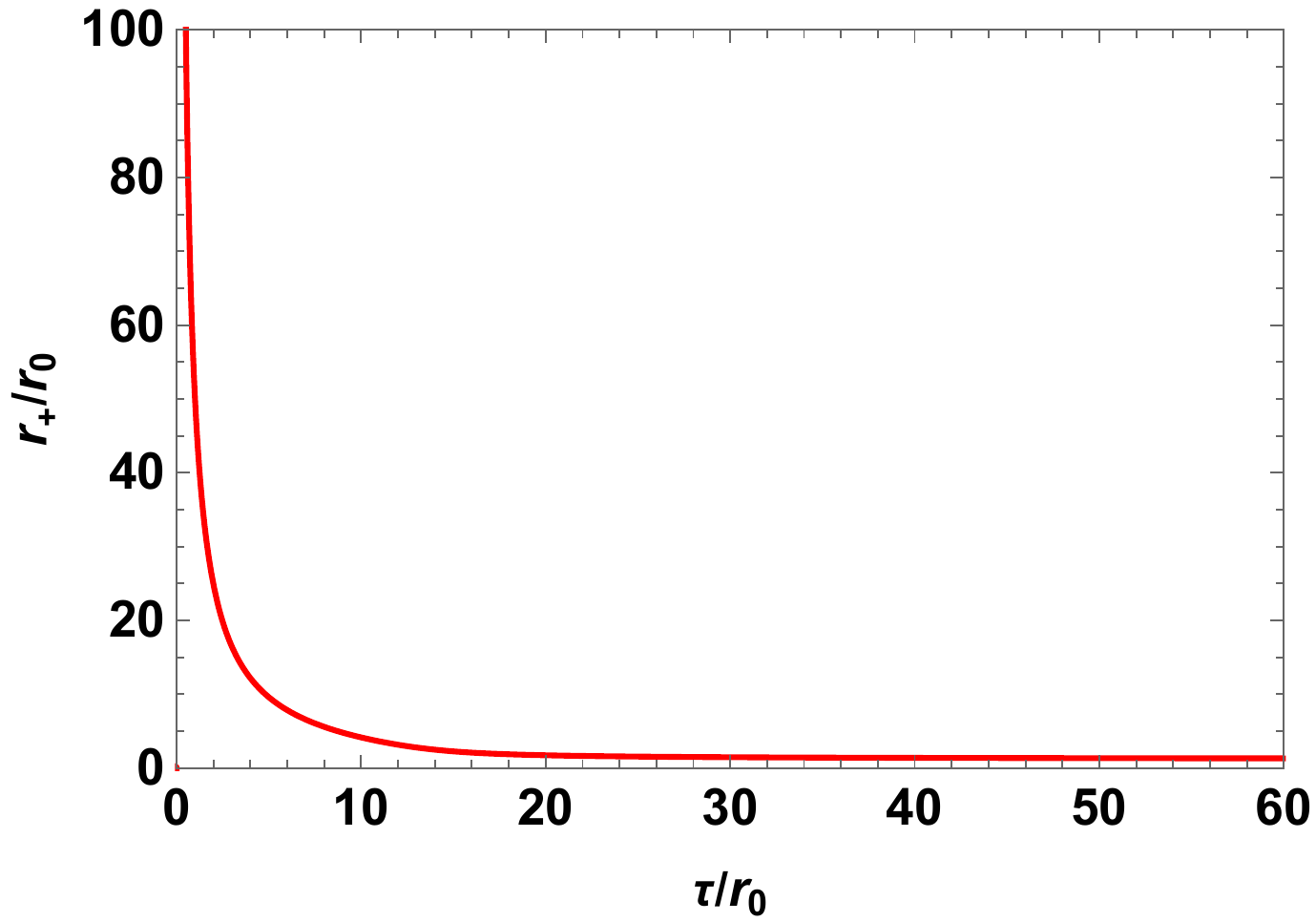}
			\caption{The zero points of $\phi^{r_+}$ in $\tau/r_0$ vs $r_+/r_0$ plane for dyonic AdS black hole in mixed ensemble for pressure greater than the critical pressure $P_c$}
			\label{Fig:Dyonic_Canonical_Thermodynamic_Defect_Tau_Plot_Above_Critical_Pressure}
		\end{subfigure} \\
	\caption{Plot of unit vector $n=(n^1,n^2)$ and zero point of $\phi^{r_+}$ for pressure $Pr_0^2=0.01$ (above the critical pressure $P_c$).}
	\label{Fig:Unit vector and zero point dyonic canonical above Pc}
	\end{figure}

For a value of pressure, $Pr_0^2=0.01$, which is above the critical pressure $P_c$,  and  $q_e/r_0=q_m/r_0=1$,   the plot of $r_+$ vs $\tau$  is shown in \autoref{Fig:Dyonic_Canonical_Thermodynamic_Defect_Tau_Plot_Above_Critical_Pressure}. In this case,  the plot exhibits only one branch corresponding to stable black hole region with positive specific heat. The winding numbers of zero points on this branch is computed to be $w=+1$. The topological number is, hence, $W=+1$. Notably, we don't find any generation/annihilation point in this case.  The zero point for $\tau/r_0=30$, $q_m/r_0=q_e/r_0=1, Pr_0^2=0.01$  is shown in \autoref{Fig:Dyonic_Canonical_Thermodynamic_Defect_Zero_Point_Vector_Above_Critical_Pressure}. What we have seen is that the topological number of $4d$ dyonic AdS black hole in canonical ensemble is not altered by a variation in pressure.

\begin{figure}[h!]
	\centerline{
	\includegraphics[scale=0.85]{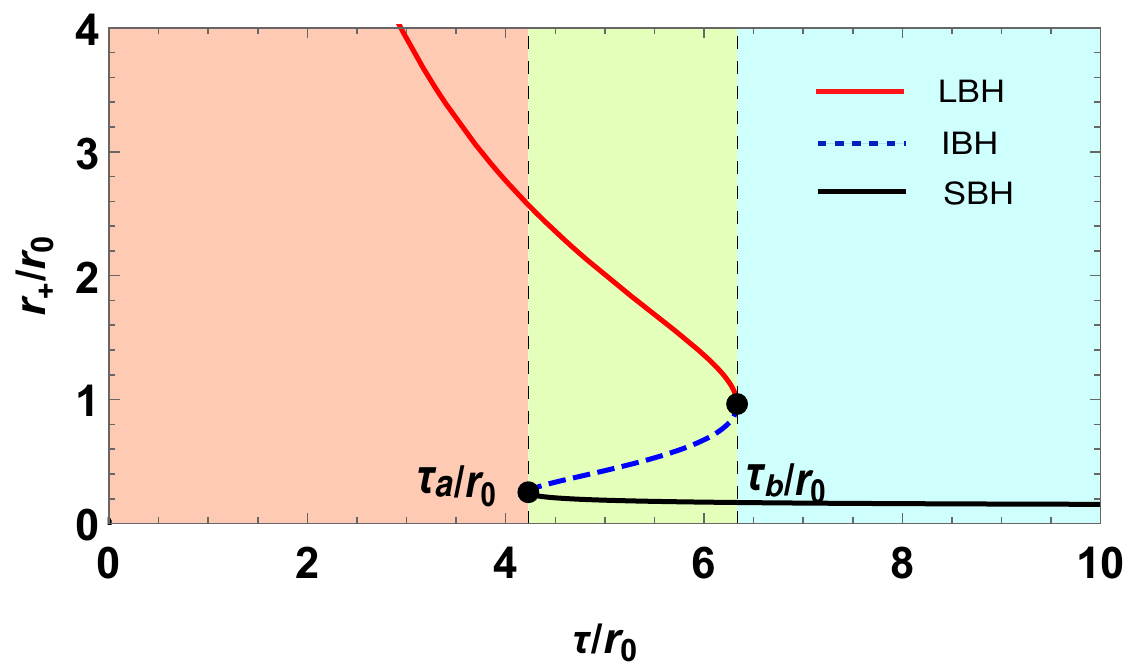}}
	\caption	{ The zero points of $\phi^{r_+}$ in $\tau/r_0$ vs $r_+/r_0$ plane for dyonic AdS black hole in canonical ensemble for $q_e/r_0=0.1$, $q_m/r_0=0.1$ and $Pr_0^2=0.04$.}
	\label{Fig:Dyonic_Canonical_fixed_P_Qm_Qe_Variation_Tau}
	\end{figure}

We repeated our analysis by changing the values of $q_e$ and $q_m$. We found that the topological number was always equal to $(W=+1)$ for all the charge configurations. As an example, the $r_+$ vs $\tau$ curve) for $q_e/r_0=0.1$, $q_m/r_0=0.1$ and $Pr_0^2=0.04$ is shown in \autoref{Fig:Dyonic_Canonical_fixed_P_Qm_Qe_Variation_Tau}. The topological number of $4d$ dyonic AdS black hole in canonical ensemble is not influenced by a variation in charge configuration.

\section{Dyonic AdS black hole in mixed ensemble}
\label{Section:Dyonic_Mixed_Canonical}
In mixed ensemble, the electric potential $\phi_e$ and the magnetic charge $q_m$  are kept constant. The electric potential $\phi_e$ is defined as 
\begin{equation}
\phi_e=\frac{q_e}{r_+},
\end{equation}
The mass and temperature, are, then modified as
\begin{equation}
M=\frac{3 r_+^2 \phi _e^2+3 q_m^2+8 \pi  P r_+^4+3 r_+^2}{6 r_+},
\end{equation}
and
\begin{equation}
\label{Eq:Dyonic_mixec_canonical_Tem_general_exp}
T=\frac{8 \pi  P r_+^4+r_+^2-r_+^2 \phi _e^2-q_m^2}{4 \pi  r_+^3}.
\end{equation}

\subsection{Topology of dyonic AdS black hole in mixed ensemble}
\begin{figure}[h!]
	\centerline{
	\includegraphics[scale=0.6]{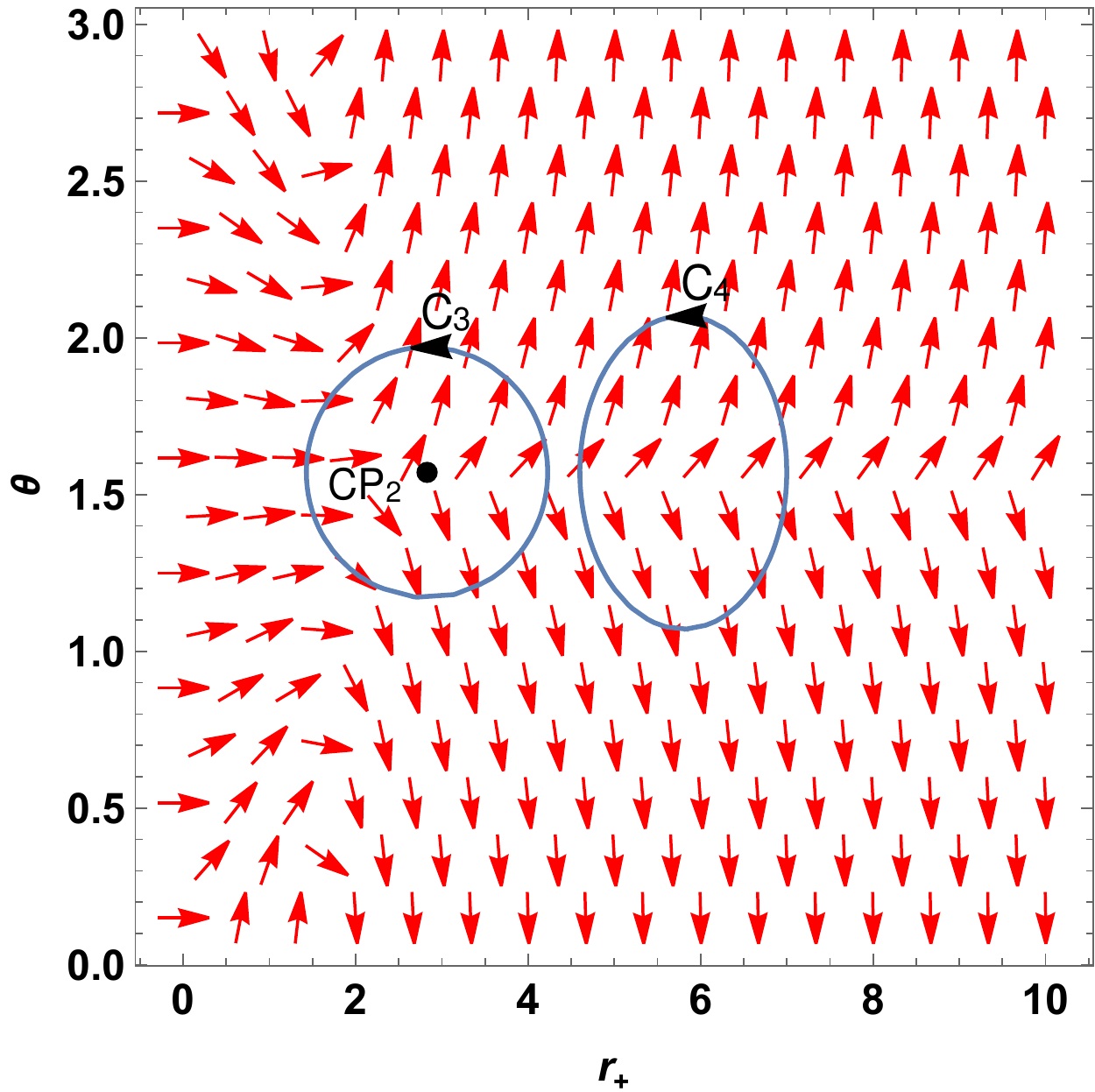}}
	\caption	{Plot of the normalized vector field $n$ in $r_+$ vs $\theta$ plane for dyonic AdS black hole in mixed ensemble. The black dot represents the critical point.}
	\label{Fig:Dyonic_Mixed_Canonical_Topology_Vector_Field}
	\end{figure}

In this section, we study the thermodynamic topology of $4d$ dyonic AdS black hole in mixed ensemble. We begin by eliminating pressure from \eqref{Eq:Dyonic_mixec_canonical_Tem_general_exp}
which is then simplified to
\begin{equation}
T(\phi_e,q_m,r_+)=-\frac{r_+^2 \left(\phi _e^2-1\right)+2 q_m^2}{2 \pi  r_+^3},
\label{Eq:Temperature_Mixed_Canonical}
\end{equation}
and the thermodynamic function $\Phi$ becomes
\begin{equation}
	\begin{aligned}
		\Phi&=\frac{1}{\sin \theta}T(\phi_e,q_m,r_+)
			&=-\frac{\csc \theta  \left\{r_+^2 \left(\phi _e^2-1\right)+2 q_m^2\right\}}{2 \pi  r_+^3}
	\end{aligned}
\end{equation}
The vector components of $\phi=( \phi^{r_+},\phi^\theta )$ are given by,
\begin{equation}
\phi^{r_+}=\frac{\csc \theta  \left\{ r_+^2 \left(\phi _e^2-1\right)+6 q_m^2\right\} }{2 \pi  r_+^4},
\end{equation}
and
\begin{equation}
\phi^{\theta}=\frac{\cot \theta  \csc \theta  \left\{ r_+^2 \left(\phi _e^2-1\right)+2 q_m^2\right\} }{2 \pi  r_+^3}
\end{equation}
The vector $\phi$ is normalized and the components are 
\begin{equation}
\label{Eq:Mixed_Canonical_Normalizd_Vector_Component_1}
\frac{\phi^{r_+}}{||\phi||}=\frac{r_+^2 \left(\phi _e^2-1\right)+6 q_m^2}{\sqrt{r_+^2 \cot ^2 \theta  \left\{ r_+^2 \left(\phi _e^2-1\right)+2 q_m^2\right\}{}^2+\left\{ r_+^2 \left(\phi _e^2-1\right)+6 q_m^2\right\} {}^2}},
\end{equation}
and
\begin{equation}
\frac{\phi^\theta}{||\phi||}=\frac{r_+ \cot \theta  \left(r_+^2 \left(\phi _e^2-1\right)+2 q_m^2\right)}{\sqrt{r_+^2 \cot ^2 \theta  \left\{ r_+^2 \left(\phi _e^2-1\right)+2 q_m^2\right\} {}^2+\left\{ r_+^2 \left(\phi _e^2-1\right)+6 q_m^2\right\} {}^2}}.
\end{equation}
Now, we plot the normalized vector $n=\Big(\frac{\phi^{r_+}}{||\phi||},\frac{\phi^{\theta}}{||\phi||}\Big)$ in $r_+$ vs $\theta$ plane by fixing $q_m=1$ and $\phi_e=1/2$  (see \autoref{Fig:Dyonic_Mixed_Canonical_Topology_Vector_Field}). Here, we find a single critical point $CP_2$ at $(r_+,\theta)=(\sqrt{6} q_m/\sqrt{1-\phi _e^2},\pi/2)$ represented by the black dot. 

\begin{figure}[h!]
	\centerline{
	\includegraphics[scale=0.7]{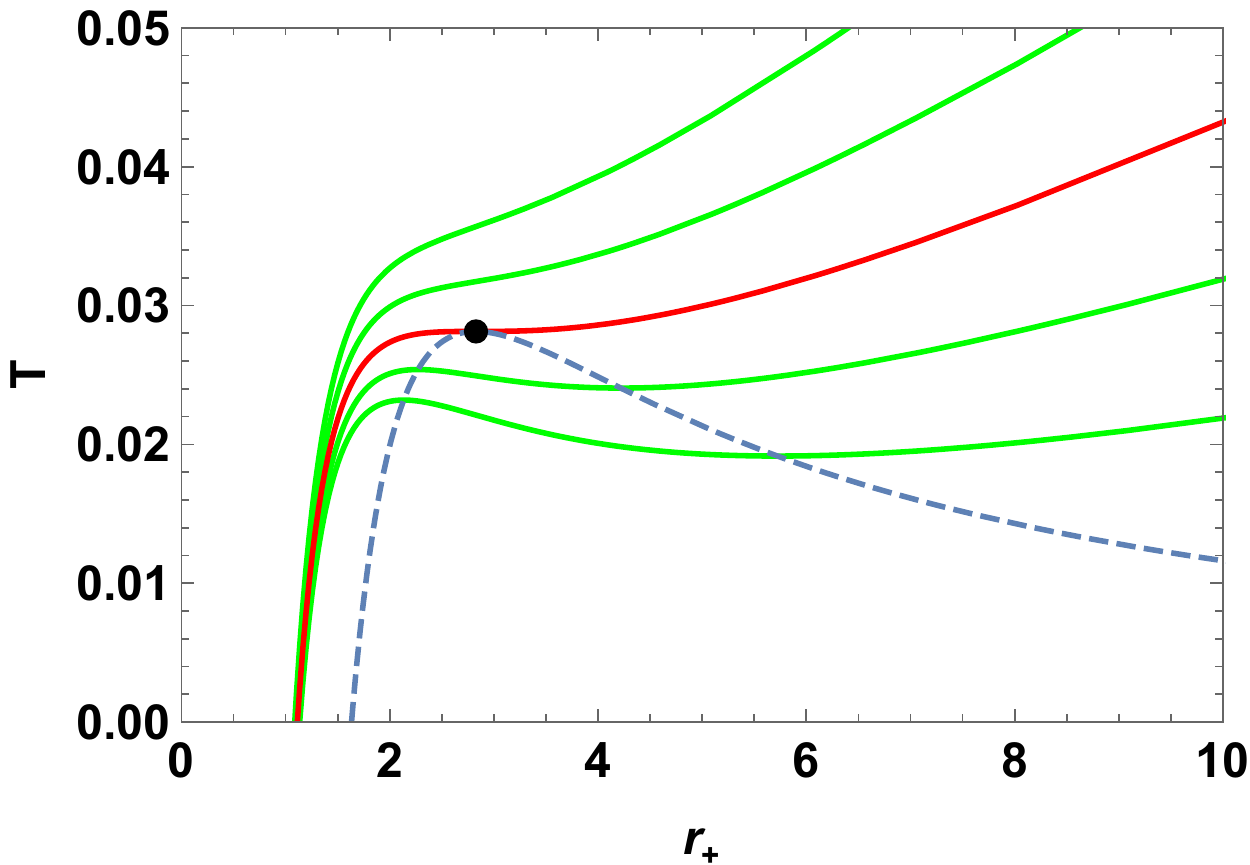}}
	\caption	{Isobaric curves (red and green) of dyonic AdS black hole in mixed ensemble. Red curve is the isobaric curve for $P=P_c$. The isobaric curves above and below the red curve is for $P>P_c$ and $P<P_c$ respectively. Black dot represents the critical point.}
	\label{Fig:Dyonic_Mixed_Canonical_Critical_Point_Isobaric_Curve}
	\end{figure}

We draw two contours $C_3$ and $C_4$ for $(a,b,r_0)=(1.4,0.4,2\sqrt{2})$ and $( 1.2,0.5,5.8)$. The contour $C_3$ encloses the critical point $CP_2$ whereas $C_4$ does not enclose any critical point. The topological charge corresponding to the contour $C_3$ is $-1$ which implies that it is a conventional critical point. The contour $C_4$ does not enclose any critical point and hence  the topological charge is $0$. The total topological charge is $-1$.
\begin{figure}[h!]
	\centerline{
	\includegraphics[scale=0.7]{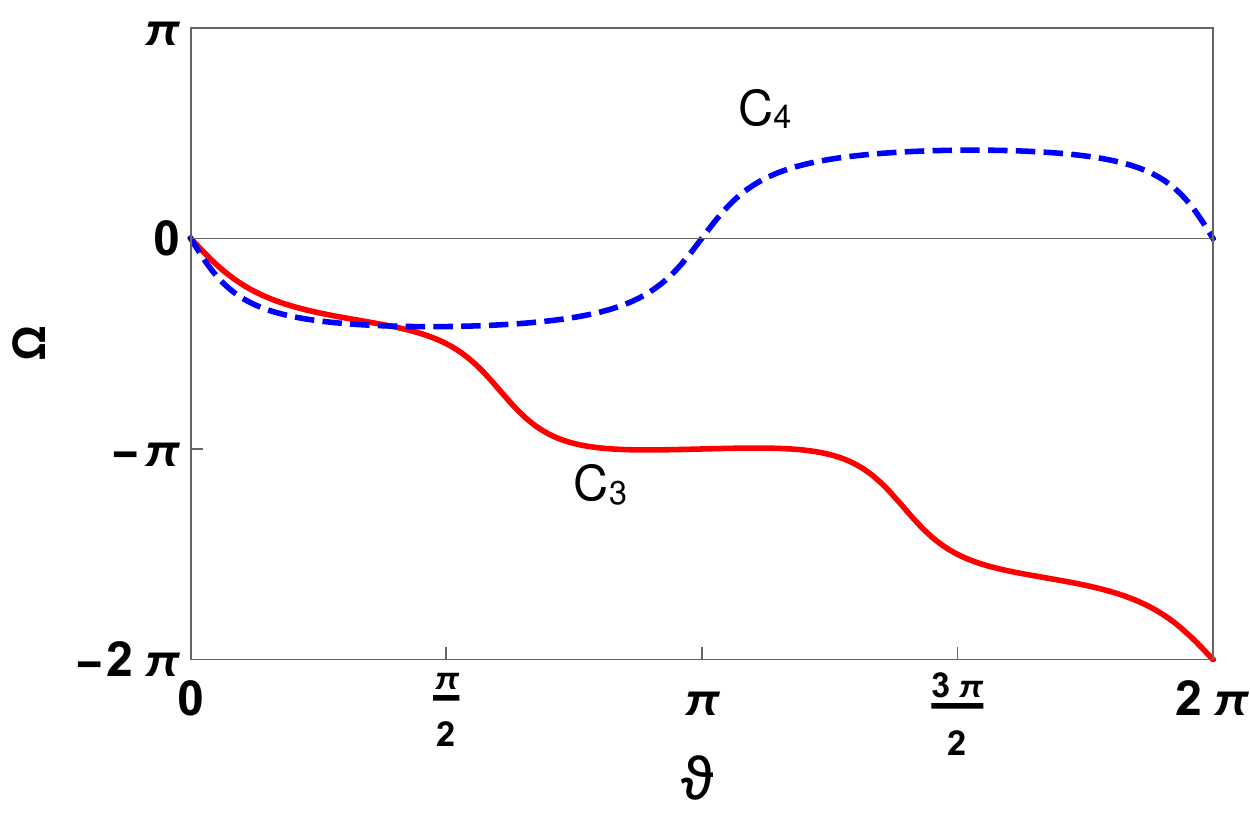}}
	\caption	{$\Omega$ vs $\vartheta$ plot for the contour $C3$ and $C_4$}
	\label{Fig:Dyonic_Mixed_Canonical_Topology_Deflection_Angle}
	\end{figure}
The deflection along the contour $C_3$ and $C_4$ is shown in \autoref{Fig:Dyonic_Mixed_Canonical_Topology_Deflection_Angle}.
The equation of state for the black hole in mixed ensemble is given by,
\begin{equation}
T=\frac{8 \pi  P r_+^4+r_+^2-r_+^2 \phi _e^2-q_m^2}{4 \pi  r_+^3}.
\end{equation}
The critical values are
\begin{equation}
\label{Eq:Dyonic_Mixed_Canonical_Critical_Values}
T_c=\frac{\left(1-\phi _e^2\right){}^{3/2}}{3 \sqrt{6} \pi  q_m}, \quad P_c=\frac{\left(\phi _e^2-1\right){}^2}{96 \pi  q_m^2} \quad \text{and} \quad r_c=\frac{\sqrt{6} q_m}{\sqrt{1-\phi _e^2}}.
\end{equation}
In this ensemble also, we see that the critical radius given in \eqref{Eq:Dyonic_Mixed_Canonical_Critical_Values} is exactly the same as the conventional critical point which is $(r_+,\theta)=(\sqrt{6} q_m/\sqrt{1-\phi _e^2},\pi/2)$. We plot the $T$ vs $r_+$ isobaric curve around the critical point in \autoref{Fig:Dyonic_Mixed_Canonical_Critical_Point_Isobaric_Curve}.
This figure shows that the the critical point (black dot) is on the isobaric curve for $P=P_c$ (red curve). The blue curve gives the extremal points of the isobaric curves and is plotted using \eqref{Eq:Temperature_Mixed_Canonical}. Similar to the canonical ensemble case, the number of phases clearly decreases with the increase of pressure $P$ and disappears at the critical point $CP_3$. This implies that the critical point $CP_3$ is a phase annihilation point.

\subsection{Dyonic AdS black hole solution as topological thermodynamic defects in mixed ensemble}

\begin{figure}[ht]
	\centering
		\begin{subfigure}{0.5\textwidth}
			\centering
			\includegraphics[width=0.9\linewidth]{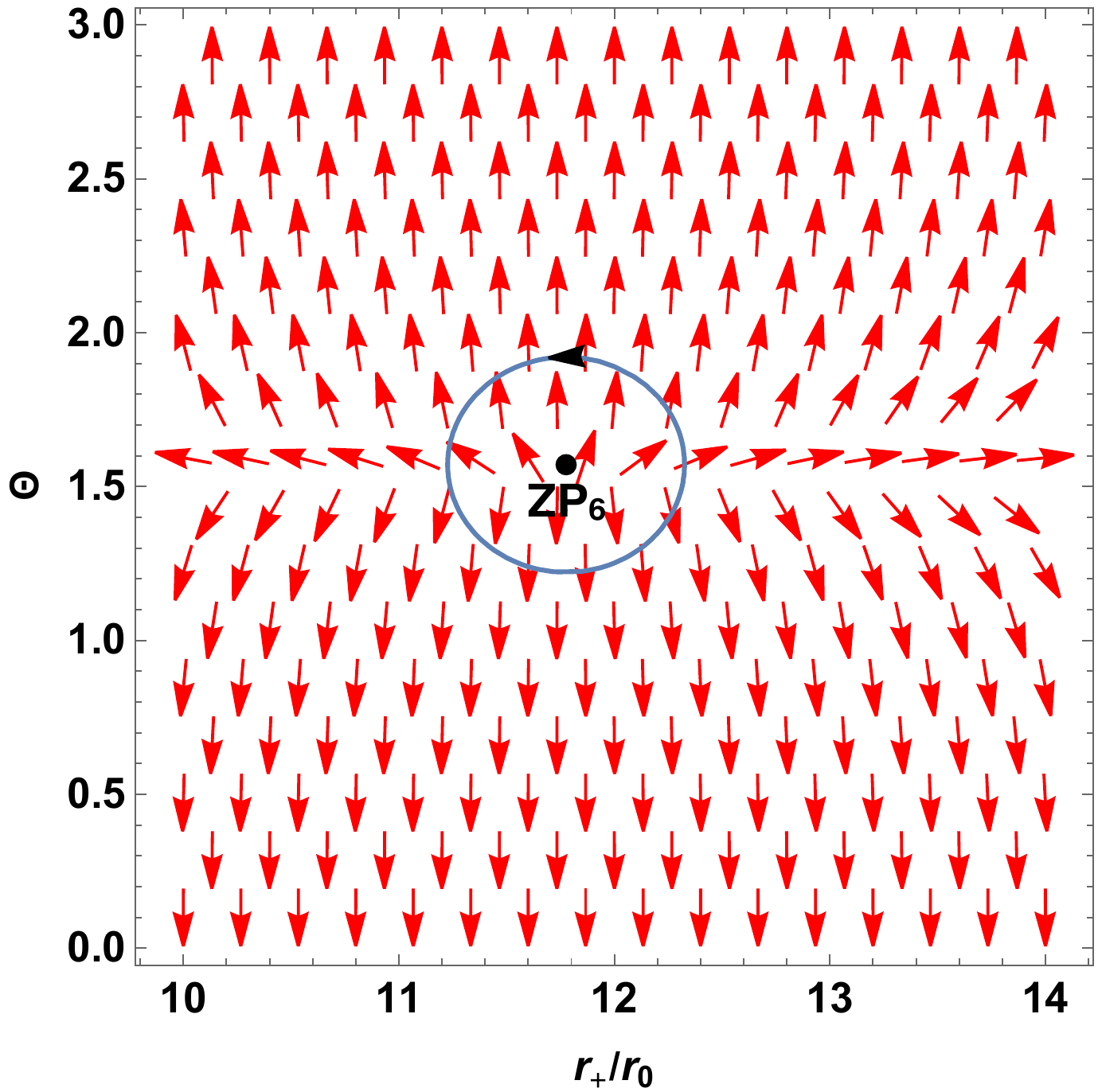}
			\caption{Unit vector $n$ in $\Theta$ vs $r_+/r_0$ plane for $\tau/r_0=35$}
			\label{Fig:Dyonic_Mixed_Canonical_Topologycal_Defect_Vector_Plot_Left_Region}
		\end{subfigure}%
		\begin{subfigure}{0.5\textwidth}
			\centering
			\includegraphics[width=0.9\linewidth]{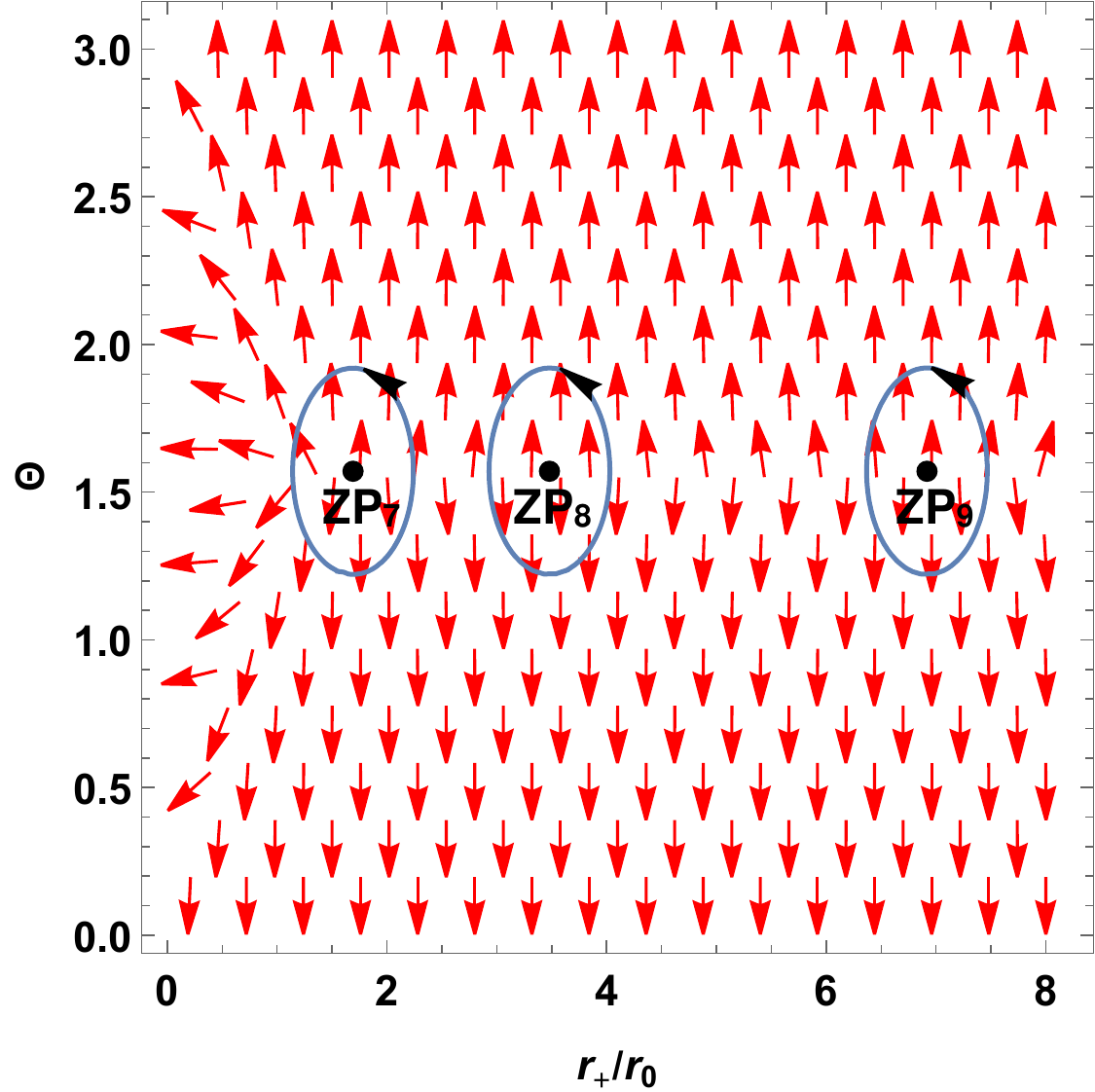}
			\caption{Unit vector $n$ in $\Theta$ vs $r_+/r_0$ plane for $\tau/r_0=45$.}
			\label{Fig:Dyonic_Mixed_Canonical_Topologycal_Defect_Vector_Plot_Middle_Region}
		\end{subfigure} 
		\begin{subfigure}{0.5\textwidth}
			\centering
			\includegraphics[width=0.9\linewidth]{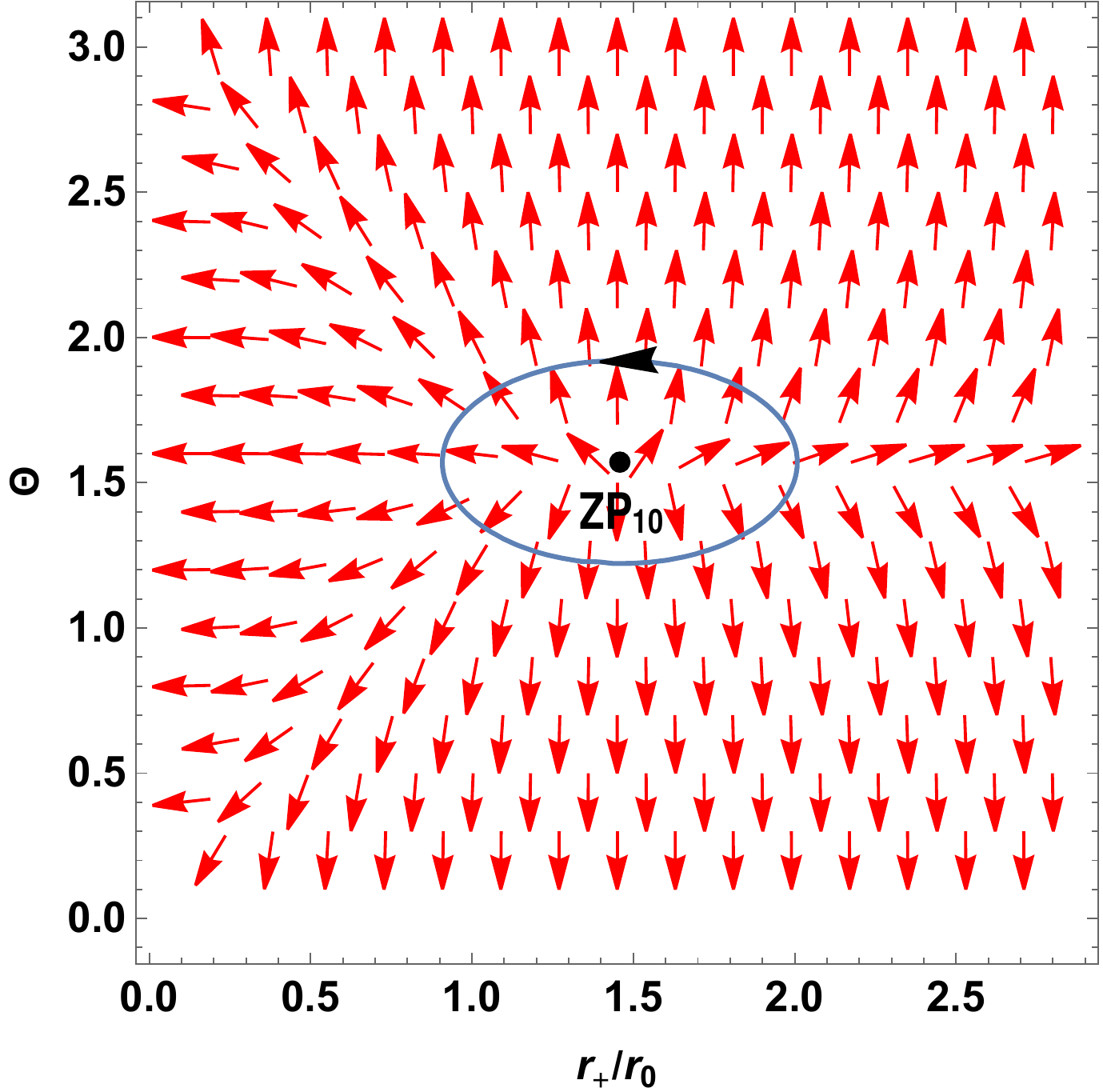}
			\caption{Unit vector $n$ in $\Theta$ vs $r_+/r_0$ plane for $\tau/r_0=55$}
			\label{Fig:Dyonic_Mixed_Canonical_Topologycal_Defect_Vector_Plot_Right_Region}
		\end{subfigure}%
	\caption{Unit vector $n=(n^1,n^2)$ shown in $\Theta$ vs $r_+/r_0$ plane for $Pr_0^2=0.001$ (pressure point below the critical pressure $P_c$). The black dots represent the zero points.}
	\label{Fig:Dyonic_Mixed_Canonical_Topologycal_Defect_Vector_Plot_All_Regions}
	\end{figure}

We now study the dyonic AdS black hole in mixed ensemble as a topological defect. As usual,  we start with the generalized free energy potential 
\begin{equation}
\label{Eq:Generalized_Free_Energy_Mixed_Canonical}
\mathcal{F}=E-\frac{S}{\tau}-q_e\phi_e.
\end{equation}
In this ensemble
\begin{equation}
E=\frac{3 r_+^2 \phi _e^2+3 q_m^2+8 \pi  P r_+^4+3 r_+^2}{6 r_+}, \quad S=\pi  r_+^2 \quad \text{and} \quad q_e=r_+\phi_e
\end{equation}
Hence,  \eqref{Eq:Generalized_Free_Energy_Mixed_Canonical} gives
\begin{equation}
\mathcal{F}=\frac{3 r_+^2 \phi _e^2+3 q_m^2+8 \pi  P r_+^4+3 r_+^2}{6 r_+}-r_+ \phi _e^2-\frac{\pi  r_+^2}{\tau }.
\end{equation}
From \eqref{Eq:Topological_Defect_Vector_Field}, the vector components can easily be worked out resulting
\begin{equation}
\phi^{r_+}=\frac{6 r_+ \phi _e^2+32 \pi  P r_+^3+6 r_+}{6 r_+}-\frac{3 r_+^2 \phi _e^2+3 q_m^2+8 \pi  P r_+^4+3 r_+^2}{6 r_+^2}-\phi _e^2-\frac{2 \pi  r_+}{\tau },
\end{equation}
and 
\begin{equation}
\phi^\Theta=-\cot \Theta  \csc \Theta .
\end{equation}
We plot the normalized vector and figure out the zero points.  The corresponding winding numbers are also calculated. For different values of $\tau/r_0$ the zero points are shown in \autoref{Fig:Dyonic_Mixed_Canonical_Topologycal_Defect_Vector_Plot_All_Regions}.
Here, we have set $\phi_e=1/2$, $q_m/r_0=1$, $Pr_0^2=0.001$ (pressure value below the critical pressure $P_c$). While for $\tau/r_0= 35$, we find one zero point $ZP_6$ with winding number $+1$, for $\tau/r_0= 45$, we locate three zero points  $ZP_7$, $ZP_8$ and $ZP_{9}$ with winding numbers $+1,-1$ and $+1$ respectively. For $\tau/r_0= 55$ , we again encounter one zero point $ZP_10$  with winding number $+1$.

\begin{figure}[h!]
	\centerline{
	\includegraphics[scale=0.7]{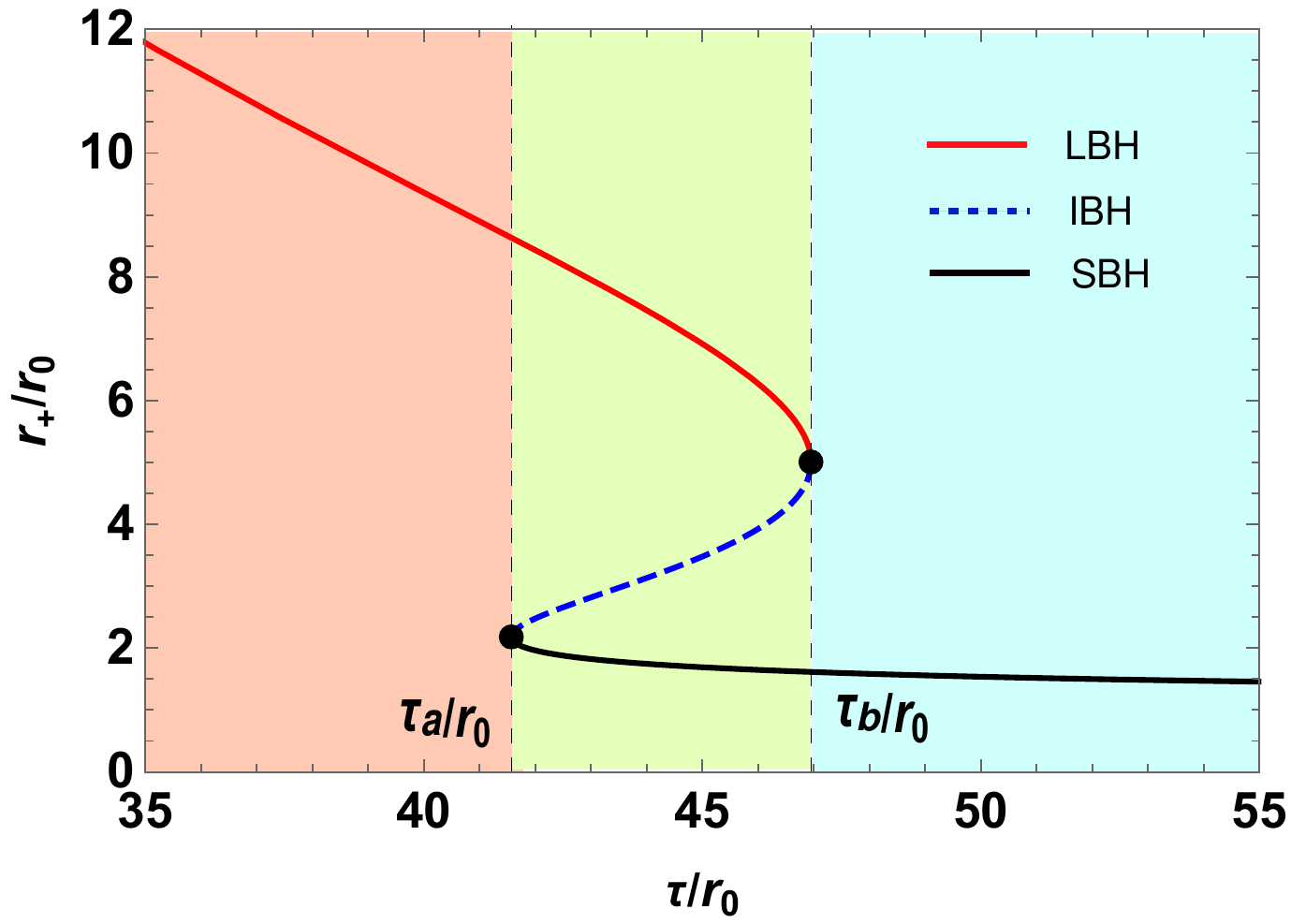}}
	\caption	{The zero points of $\phi^{r_+}$ in $\tau/r_0$ vs $r_+/r_0$ plane for dyonic AdS black hole in mixed ensemble below the critical pressure $P_c$.}	\label{Fig:Dyonic_Mixed_Canonical_Topologycal_Defect_Tau_Curve}
	\end{figure}

 The zero points of the component $\phi^{r_+}=0$ are given by the equation
\begin{equation}
\tau=\frac{4 \pi  r_+^3}{8 \pi  P r_+^4+r_+^2-r_+^2 \phi _e^2-q_m^2}.
\end{equation}
For $\phi_e=1/2$, $q_m/r_0=1$, $Pr_0^2=0.001$ the resulting $r_+$ vs $\tau$  graph is plotted in \autoref{Fig:Dyonic_Mixed_Canonical_Topologycal_Defect_Tau_Curve}.
Similar to the canonical ensemble case, below critical pressure, we have three branches of the $\tau$ curve in the regions for $\tau<\tau_b$, $\tau_a<\tau<\tau_b$ and $\tau>\tau_a$. The first and third branch are the large black hole and small black hole region and the zero points on these two branches have $w=+1$ and positive specific heat. The other branch represents intermediate black hole region and the zero points in this region have $w=-1$ and negative specific heat. Thius, the topological number is $W=+1-1+1=+1$. The generation and annihilation points for $\phi_e=1/2$, $q_m/r_0=1$, $Pr_0^2=0.001$ is found at $\tau/r_0=\tau_a/r_0=41.5688$ and $\tau/r_0=\tau_b/r_0=46.9484$ respectively. 

\begin{figure}[ht]
	\centering
		\begin{subfigure}{0.5\textwidth}
			\centering
			\includegraphics[width=0.8\linewidth]{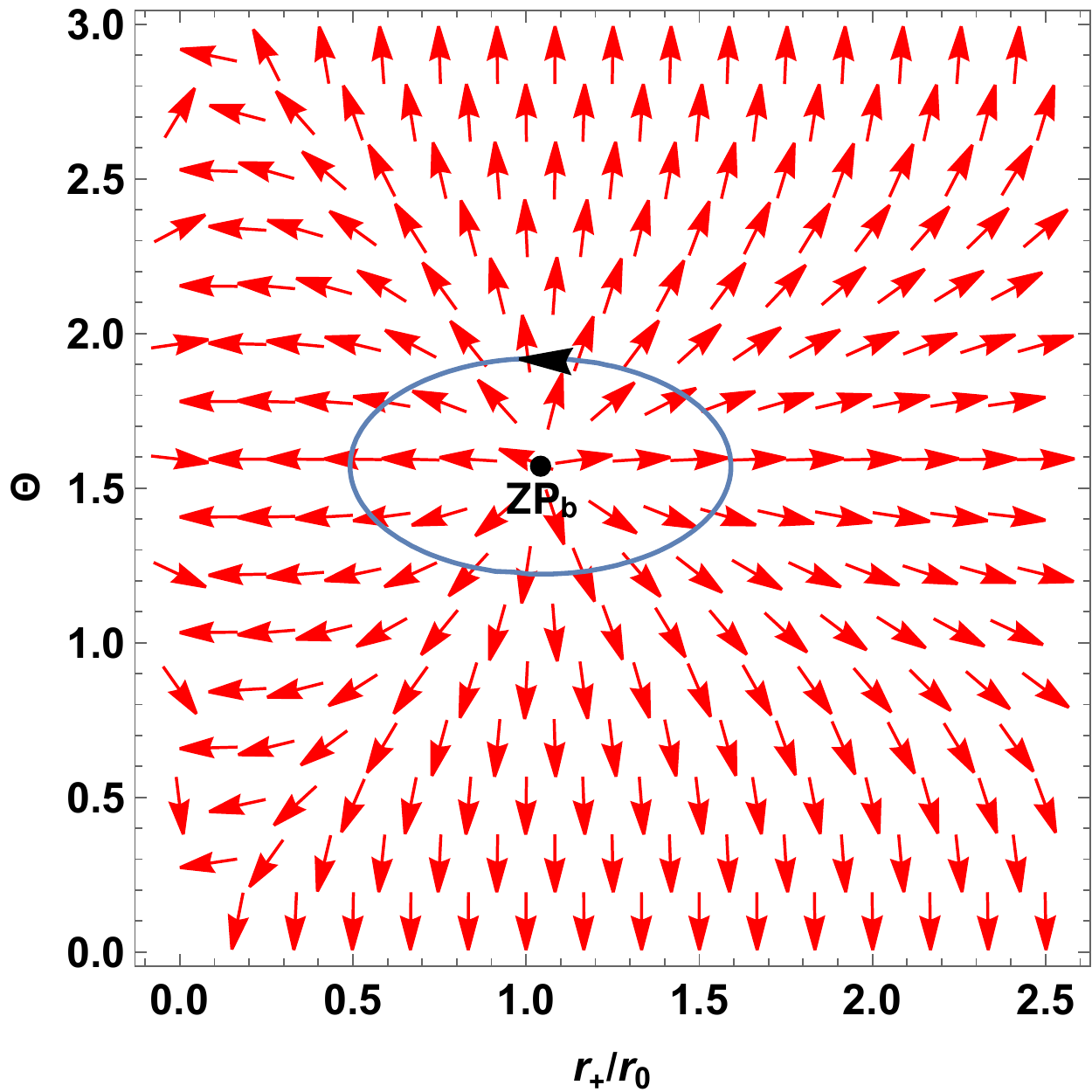}
			\caption{Unit vector $n=(n^1,n^2)$ shown in $\Theta$ vs $r_+/r_0$ plane for $Pr_0^2=0.02$. The black dot represents zero point.}
			\label{Fig:Dyonic_Mixed_Canonical_Thermodynamic_Defect_Zero_Point_Vector_Above_Critical_Pressure}
		\end{subfigure}%
		\begin{subfigure}{0.5\textwidth}
			\centering
			\includegraphics[width=1\linewidth]{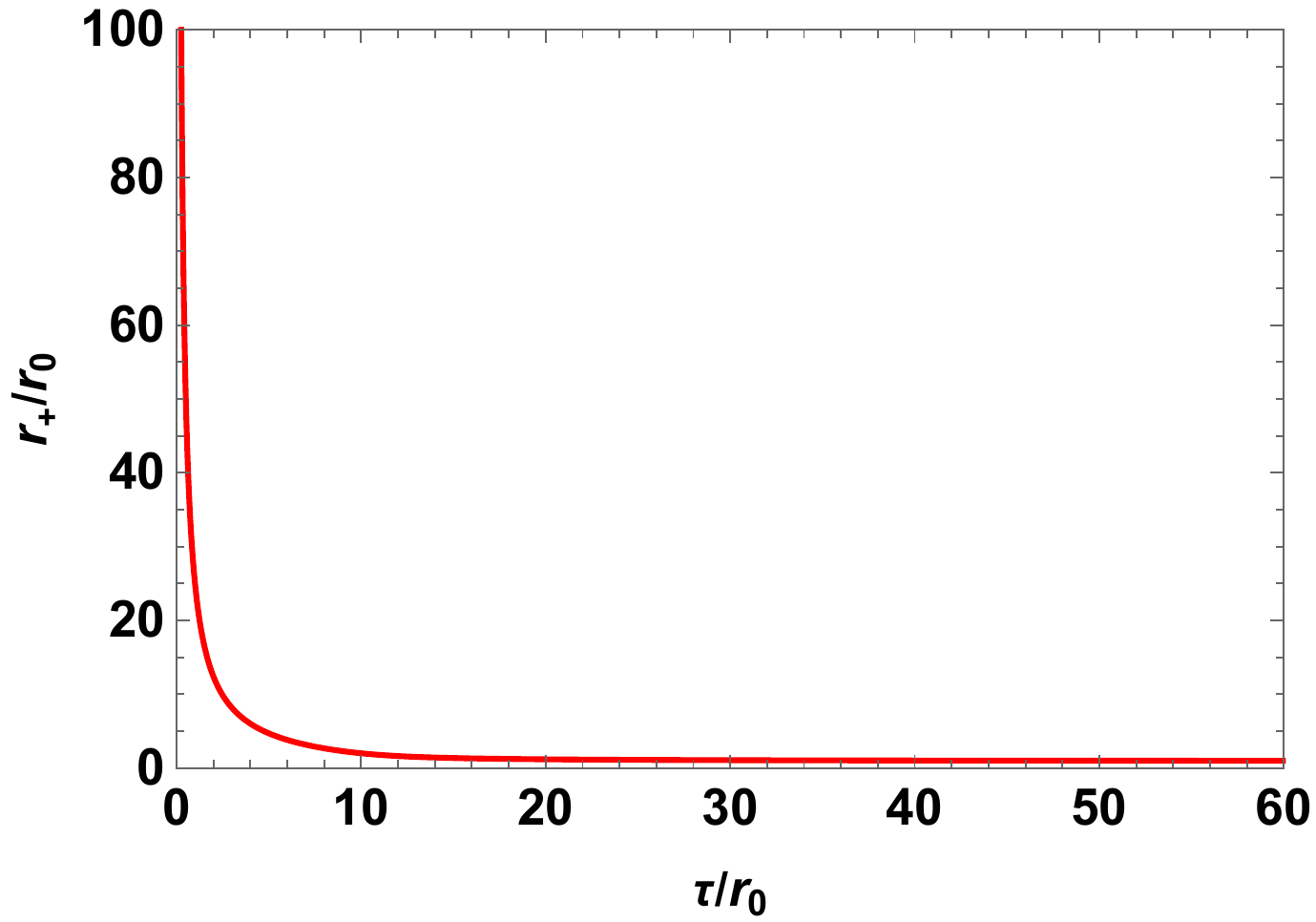}
			\caption{The zero points of $\phi^{r_+}$ in $\tau/r_0$ vs $r_+/r_0$ plane for Dyonic AdS black hole in mixed ensemble for pressure greater than the critical pressure $P_c$.}
			\label{Fig:Dyonic_Mixed_Canonical_Thermodynamic_Defect_Tau_Plot_Above_Critical_Pressure}
		\end{subfigure} \\
	\caption{Plot of unit vector $n=(n^1,n^2)$ and zero point of $\phi^{r_+}$ for pressure $Pr_0^2=0.02$ (above the critical pressure $P_c$).}
	\label{Fig:Unit vector and zero point dyonic mixed canonical above Pc}
	\end{figure}

Above critical pressure $P_c$, the same plot is shown in \autoref{Fig:Dyonic_Mixed_Canonical_Thermodynamic_Defect_Tau_Plot_Above_Critical_Pressure}.
Here, we chose $Pr_0^2=0.02$, $q_m=1$ and $\phi_e=1/2$. In this case, the unstable region disappears and the $\tau(r_+)$ curve corresponds to stable black hole region. The winding number for each point on the curve is $w=+1$ and the topological number is hence $W=+1$. The unit vector and the zero point for $\tau/r_0=35$, $Pr_0^2=0.02$, $q_m/r_0=1$ and $\phi_e=1/2$ is shown in \autoref{Fig:Dyonic_Mixed_Canonical_Thermodynamic_Defect_Zero_Point_Vector_Above_Critical_Pressure}.

\begin{figure}[h!]
	\centerline{
	\includegraphics[scale=0.85]{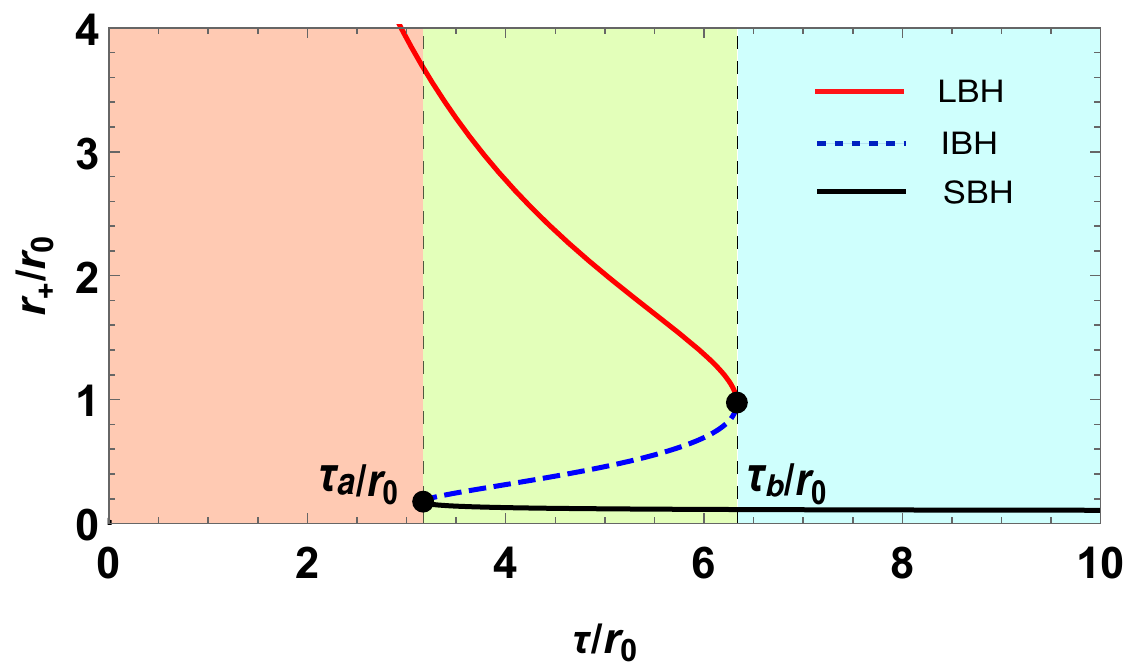}}
	\caption	{The zero points of $\phi^{r_+}$ in $\tau/r_0$ vs $r_+/r_0$ plane for dyonic AdS black hole in mixed ensemble for $\phi_e=0.1$, $q_m/r_0=0.1$ and $Pr_0^2 =0.04$.}	\label{Fig:Dyonic_Mixed_Canonical_fixed_P_Qm_Phie_Variation_Tau}
	\end{figure}

We repeated the exercise altering the values of $\phi_e$ and $q_m/r_0$ and observed that the topological number for all the combinations was identical and equal to  $W=+1$. The zero points of $\phi^{r_+}$ is shown in \autoref{Fig:Dyonic_Mixed_Canonical_fixed_P_Qm_Phie_Variation_Tau} for $\phi_e=0.1$, $q_m/r_0=0.1$ and $Pr_0^2=0.04$. 	

\section{Dyonic AdS black hole in grand canonical ensemble}
\label{Section:Dyonic_Grand_Canonical}
In the grand canonical ensemble, both the electric potential $\phi_e$  and the magnetic potential $\phi_m$ are kept fixed. 
\begin{equation}
\phi_e=\frac{q_e}{r_+}, \quad \text{and} \quad \phi_m=\frac{q_m}{r_+}.
\end{equation} 
The relevant thermodynamic parameters of dyonic AdS black hole in grand canonical ensemble are given by,
\begin{equation}
M=\frac{1}{6} r_+ \big\{3 \left(\phi _e^2+\phi _m^2+1\right)+8 \pi  P r_+^2\big\},
\end{equation}
\begin{equation}
S=\pi r^2_+, 
\end{equation}
and
\begin{equation}
T=\frac{8 \pi  P r_+^2+1- \phi _e^2-\phi _m^2}{4 \pi  r_+}.
\label{Eq:Dyonic_grand_canonical_temp_general}
\end{equation}

\subsection{Topology of dyonic AdS black hole in grand canonical ensemble}
Now, we proceed to study the topology of dyonic AdS black hole thermodynamics in grand canonical ensemble. First, we eliminate pressure from \eqref{Eq:Dyonic_grand_canonical_temp_general} 
using $\Big( \frac{\partial_{r_+} T}{\partial_{r_+}S}\Big)_{\phi_e,\phi_m,P}=0$.
\begin{equation}
T=\frac{1-\text{$\phi_e$}^2-\text{$\phi_m$}^2}{2 \pi  r_+}.
\end{equation}
The thermodynamics function $\Phi=T/\sin\theta$ is, therefore, 
\begin{equation}
\Phi=\frac{\csc \theta  \left(-\text{$\phi_e$}^2-\text{$\phi_m$}^2+1\right)}{2 \pi  r_+}.
\end{equation}
The components of vector field $\phi=( \phi^{r_+},\phi^\theta )$ are
\begin{equation}
\phi^{r_+}=\frac{\csc \theta  \left(\text{$\phi_e$}^2+\text{$\phi_m$}^2-1\right)}{2 \pi  r_+^2},
\end{equation}
and
\begin{equation}
\phi^\theta=\frac{\cot \theta  \csc \theta  \left(\text{$\phi_e$}^2+\text{$\phi_m$}^2-1\right)}{2 \pi  r_+}.
\end{equation}
The normalized vector components are
\begin{equation}
\frac{\phi^{r_+}}{||\phi||}=\frac{1}{\sqrt{r_+^2 \cot ^2(\theta )+1}},
\end{equation}
and
\begin{equation}
\frac{\phi^{\theta}}{||\phi||}=\frac{r_+ \cot (\theta )}{\sqrt{r_+^2 \cot ^2(\theta )+1}}.
\end{equation}

\begin{figure}[!ht]
	\centering
		\begin{subfigure}{0.5\textwidth}
			\centering
			\includegraphics[width=0.9\linewidth]{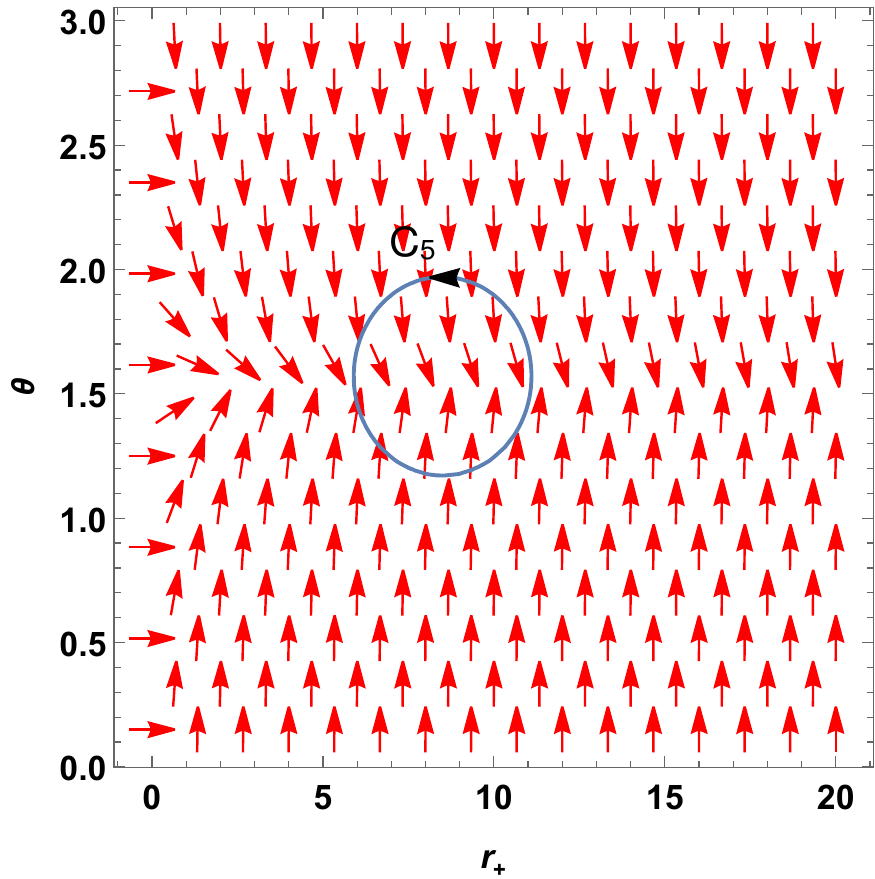}
			\caption{Plot of the normalized vector field $n$ in $r_+$ vs $\theta$ plane for Dyonic AdS black hole in grand canonical ensemble. There is no critical point.}
			\label{Fig:Dyonic_Grand_Canonical_Topology_Vector_Field}
		\end{subfigure}%
		\begin{subfigure}{0.5\textwidth}
			\centering
			\includegraphics[width=0.9\linewidth]{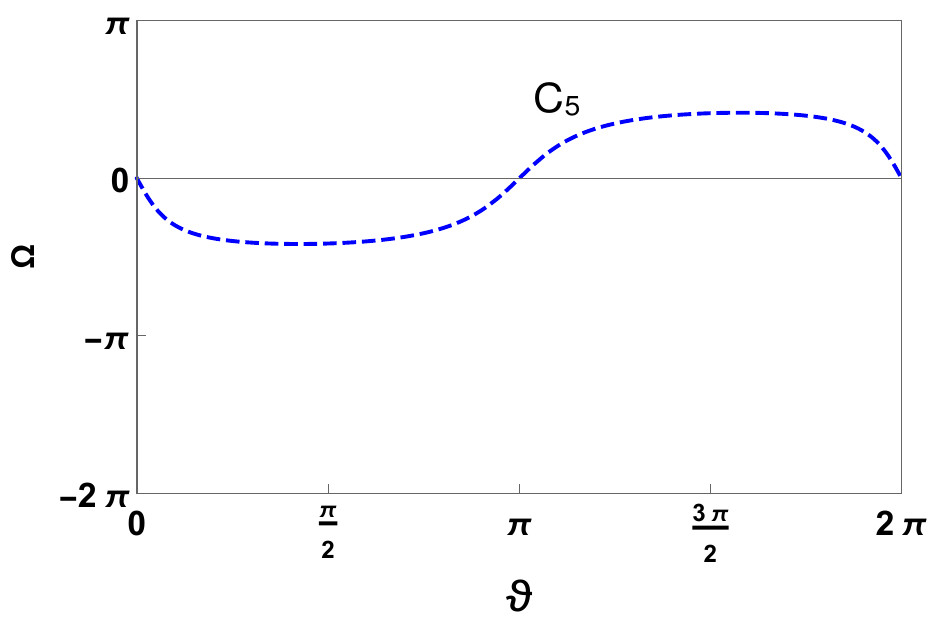}
			\caption{$\Omega$ vs $\vartheta$ plot for the contour $C5$}
			\label{Fig:Dyonic_Grand_Canonical_Topology_Deflection_Angle}
		\end{subfigure} 
	\caption{Plot of vector $n$ in $r_+$ vs $\theta$ plane and the plot of deflection angle $\Omega(\vartheta)$ along $C_5$. }
	\label{Fig:Dyonic_Grand_Canonical_Nornalized_Vector_and_Deflection_Angle}
	\end{figure}

On following the procedure discussed earlier, we find that this system does not have any critical points (see \autoref{Fig:Dyonic_Grand_Canonical_Topology_Vector_Field}). Also, since the contour does not enclose any critical point therefore, the $\Omega(\vartheta)$ function reaches $0$ at $\vartheta=2\pi$. (see \autoref{Fig:Dyonic_Grand_Canonical_Topology_Deflection_Angle})

\subsection{Dyonic AdS black hole solution as topological thermodynamic defects in grand canonical ensemble}

Now, we identify dyonic AdS black hole as a topological defect in the thermodynamic we can use the following general free energy potential: 
\begin{equation}
\label{Eq:Generalized_Free_Energy_Grand_Canonical}
\mathcal{F}=E-\frac{S}{\tau}-q_e\phi_e-q_m\phi_m.
\end{equation}
Following \eqref{Eq:Topological_Defect_Vector_Field}, we calculate the vector components 
Accordingly, we find the unit vectors
\begin{equation}
\phi^{r_+}=\frac{1}{6} \left(3 \phi _e^2+3 \phi _m^2+8 \pi  P r_+^2+3\right)-\phi _e^2-\phi _m^2+\frac{8}{3} \pi  P r_+^2-\frac{2 \pi  r_+}{\tau },
\end{equation}
and
\begin{equation}
\phi^{\Theta}=-\cot \Theta  \csc \Theta.
\end{equation}

\begin{figure}[h!]
	\centerline{
	\includegraphics[scale=0.7]{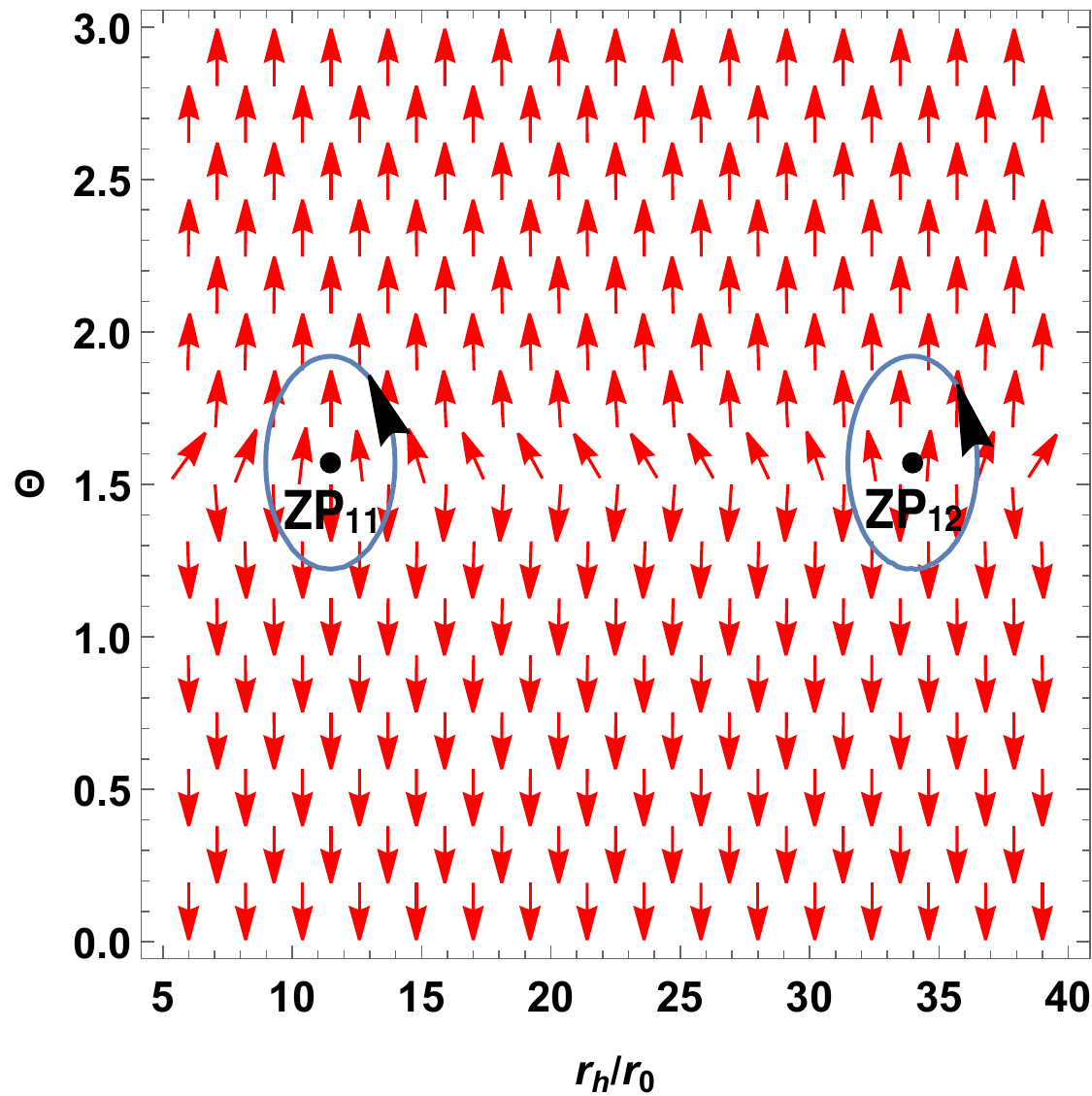}}
	\caption{Unit vector $n=(n^1,n^2)$ shown in $\Theta$ vs $r_+/r_0$ plane for $\tau/r_0= 110$,  $\phi _e=0.1$,  $\phi _m=0.1$ and $Pr_0^2=0.0001$. The black dots represent the zero points.}
	\label{Fig:Dyonic_Grand_Canonical_Topologycal_Defect_Vector_Plot}
	\end{figure}

For $\tau/r_0= 110$,  $\phi _e=0.1$,  $\phi _m=0.1$ and $Pr_0^2=0.0001$, we find two zero points $ZP_{11}$ and $ZP_{12}$ with winding numbers $-1$ and $+1$ respectively. These are shown in \autoref{Fig:Dyonic_Grand_Canonical_Topologycal_Defect_Vector_Plot}. Hence, the topological number of dyonic AdS black hole in grand canonical ensemble is $0$. 

\begin{figure}[htb!]
	\centerline{
	\includegraphics[scale=0.7]{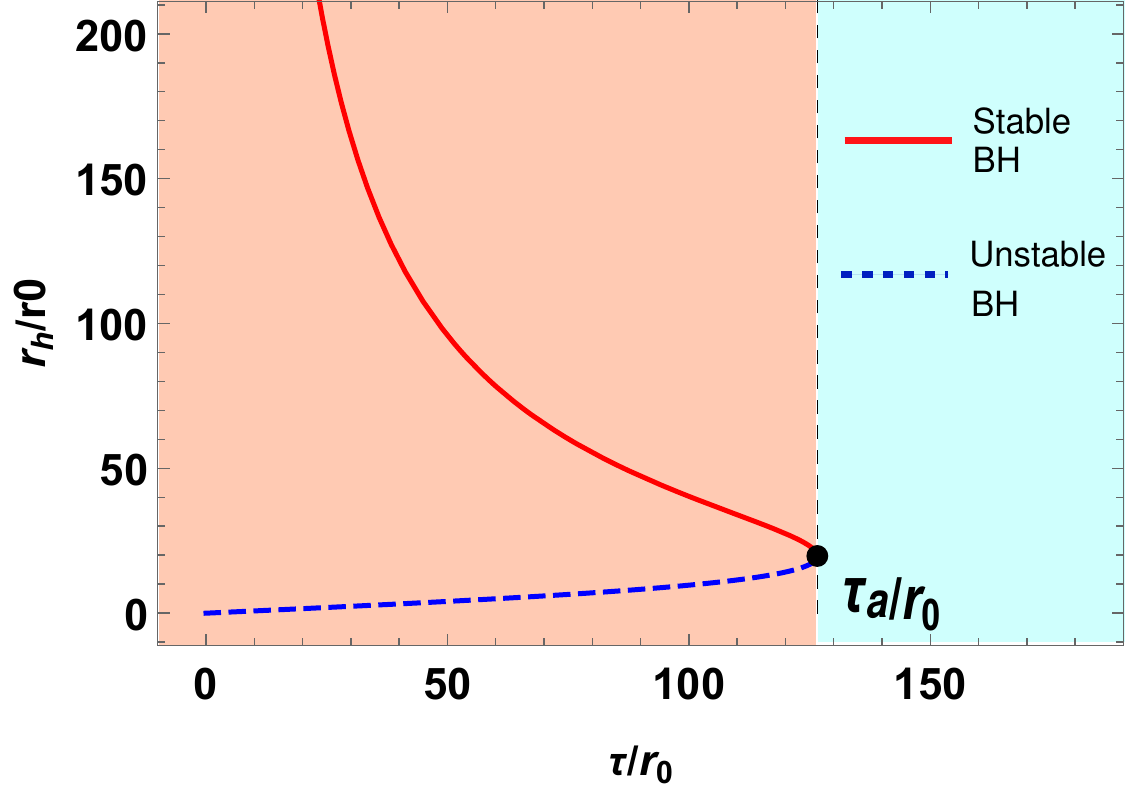}}
	\caption{The zero points of $\phi^{r_+}$ in $\tau/r_0$ vs $r_+/r_0$ plane for dyonic AdS black hole in grand canonical ensemble}
	\label{Fig:Dyonic_Grand_Canonical_Topologycal_Defect_Tau_Curve}
	\end{figure}

The expression for $\tau$ representing zero points is given by
\begin{equation}
\tau=\frac{4 \pi  r_+}{8 \pi  P r_+^2+1-\phi _e^2-\phi _m^2}
\label{Eq:Dyonic_Grand_Canonical_Ensemble_Tau}
\end{equation}
For $\phi_e=0.1$, $\phi_m=0.1$ and $Pr_0^2=0.0001$, $r_+/r_0$ vs $\tau/r_0$ plot is  shown in \autoref{Fig:Dyonic_Grand_Canonical_Topologycal_Defect_Tau_Curve}.

In this case, unlike what we saw in canonical and mixed ensembles, the plot exhibits two black hole branches in the regions $\tau<\tau_a$ and $\tau>\tau_a$. The former branch represents unstable black hole region whereas the later branch represents stable black hole region. Any zero point in the unstable and stable region has winding numbers $w=-1$ and $w=+1$  respectively. The topological number is hence $W=0$ which is different from what we had in canonical and mixed ensembles. The creation point is  located at $\tau/r_0=\tau_a/r_0=126.604 $. 

\begin{figure}[h!]
	\centerline{
	\includegraphics[scale=0.7]{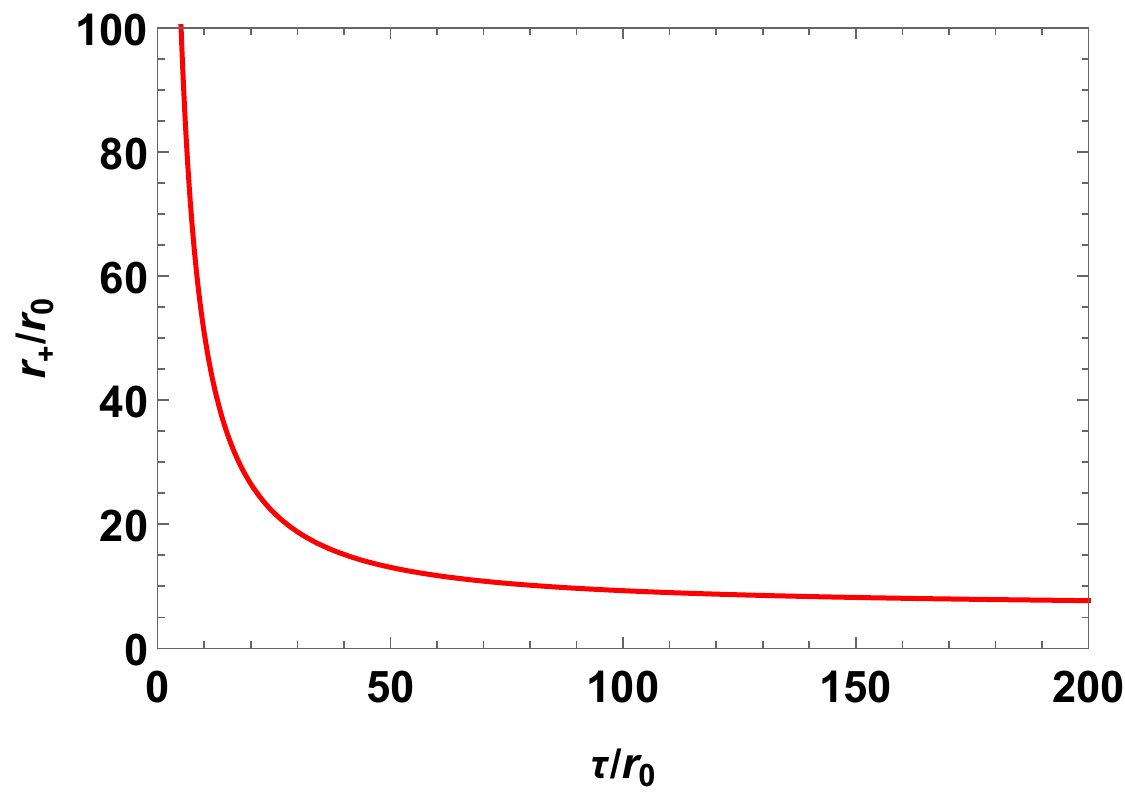}}
	\caption	{The zero points of $\phi^{r_+}$ in $\tau/r_0$ vs $r_+/r_0$ plane for dyonic AdS black hole in grand canonical ensemble for $\phi_e^2+\phi_m^2>1$ case. Here, we choose $\phi_e=1$, $\phi_m=1$ and $Pr_0^2=0.001$.}
\label{Fig:Dyonic_Grand_Canonical_Topologycal_Defect_Tau_Curve_no-unstable-region}
	\end{figure}

Interestingly, for  $\phi_e^2+\phi_m^2>1$, we get only one black hole branch with winding numbert $+1$ (see \autoref{Fig:Dyonic_Grand_Canonical_Topologycal_Defect_Tau_Curve_no-unstable-region}). The topological number, therefore, is $+1$. In this case, we do not see any generation or annihilation point.

Thus, in the grand canonical ensemble, for$4d$ dyonic AdS black hole, we have two topological numbers. When $\phi_e^2+\phi_m^2<1$, the topological number is $0$ in contrast to what we had in canonical and mixed ensembles.  When $\phi_e^2+\phi_m^2>1$, the topological number is $1$, same as the ones in canonical and mixed ensembles.
\newpage
\section{Conclusion}
In this work, we studied the thermodynamic topology of $4$d dyonic AdS black hole in canonical, mixed and grand canonical ensembles. Canonical, mixed and grand canonical ensembles were formed by fixing electric and magnetic charges,  magnetic charge and electric potential and potentials corresponding to both electric and magnetic charges respectively. In all the three ensembles, we evaluated the topological charges of the critical points in their thermodynamic spaces. We observed  the presence of a solitary conventional critical point with topological charge $-1$ in both canonical and mixed ensembles. Contrastingly, in the grand canonical ensemble, no critical point was found. Next, we recognized the dyonic AdS black hole as topological defects in thermodynamic space and analyzed its local and global topology by calculating the winding numbers at the defects. We found that, in both canonical and mixed ensembles, the total topological charge was equal to $1$, which was not altered by changes in thermodynamic parameters. In both these ensembles , either one generation and one annihilation points (below critical pressure) or no generation/annihilation points (above critical pressure) were seen. In the grand canonical ensemble, depending on the values of potentials, the total topological charge was found to be either equal to $0$ (when $\phi_e^2+\phi_m^2<1$)  or $1$  (when $\phi_e^2+\phi_m^2>1$) .  In this ensemble, we  found either one generation point (when $\phi_e^2+\phi_m^2<1$) or no generation/annihilation point ( when $\phi_e^2+\phi_m^2>1$).\\

From our analysis, we conclude that $4$d dyonic AdS black hole in canonical and mixed ensembles can be placed in the same thermodynamic topological class. However, the thermodynamic topology of $4$d dyonic AdS black hole in grand canonical ensemble is different from those in the other two ensembles. Or in other words, the topological class of $4$d dyonic AdS black hole is ensemble dependent. It will be interesting to extend the study of ensemble dependent thermodynamic topology to other black holes with rich phase structures. We plan to do so in our future works.

\label{Section:Canonical}


\appendix*

\bibliographystyle{apsrev}
\end{document}